\documentclass[useAMS,usenatbib]{mn2e}
\usepackage{graphicx}

\begin{document}

\title{A method of detecting  radio transients}
\author{P. A. Fridman$^{1}$
\thanks{
E-mail: fridman@astron.nl}
\\
$^{1}$ASTRON, Dwingeloo, Postbus 2, 7990AA, The Netherlands}

\date{Accepted . Received   ; in original form
}

\maketitle

\begin{abstract}
Radio transients are  sporadic signals and their detection requires that the  backends of radio telescopes be equipped with the appropriate hardware and  software  to undertake this.
Observational programs to detect transients can be dedicated or they can piggy-back on observations made by other programs. It is the single-dish single-transient (non-periodical) mode which is considered in this paper. Because neither the width of a transient nor the time of its arrival is  known, a sequential analysis in the form of a cumulative sum (cusum) algorithm is proposed here. Computer simulations and real observation data processing are
included to demonstrate the performance of the cusum.  The use of the Hough transform is here proposed for the purpose of non-coherent de-dispersion. It is possible that the detected transients could be radio frequency interferences (RFI) and a procedure is proposed here which can distinguish between    celestial signals and  man-made RFI. This procedure is based on an analysis of the statistical properties  of the signals.
\end{abstract}

\pagerange{\pageref{firstpage}--\pageref{lastpage}} \pubyear{2010}

\label{firstpage}


\section{Introduction}

Radio astronomy signals received by radio telescopes have  noise-like waveforms
which have a normal (Gaussian) probability distribution function (pdf) ${\cal N} (0,\sigma_{s})$, i. e., with zero mean and variance
$\sigma_{s}^{2}$. Background radio emission and radio receivers also produce normal noise ${\cal N} (0,\sigma_{sys})$.
Basically, radio astronomy observations consist of detecting and  measuring $\sigma_{s}^{2}$ at the background of $\sigma_{sys}^{2}$.  This is valid for  total power radiometry
 (spatial distribution of  signal noise power), spectrography (temporal coherence measurements)  and interferometry (spatial coherence measurements).
 Total power radiometry is involved with  the intensity variability of radio sources and, in particular, with  sporadic phenomena - transient radio emissions. The time scale of  radio transients can span  nanoseconds to days. Traditional total power radiometers are not equipped with   backends (both hardware and software) designed to detect transients. Many  transients were found in a serendipitous way and the discoveries of pulsars and RRAT are the most prominent examples of such discoveries.  Systematic searching for non-periodical transients began only recently (in the last decennia). Future radio telescopes (ATA, MWA, LOFAR, SKA) will  be able to provide more  opportunities for the detecting of transients.
 \citep{Cordes3}.

Several works dedicated to  single radio transient detection have been published. Interstellar scattering and scintillation, effective time resolution, de-dispersion methods, matched filtering and thresholding have been considered in \citep{Cordes1}.\\
 A transient surveys strategy and search processing have been studied  \citep{Cordes2}.  The importance of single-pulse detection vs. many-pulse detection for highly modulated pulse trains is demonstrated in this work and also in \citep{McLaughlin}. The influence of the amplitude probability distribution of  transients was studied in detail in this latter article.

 The search for transients consists of two main operations:

1. De-dispersion aimed at removing the effects of interstellar dispersion. There are two methods of de-dispersion:\\
a) Coherent de-dispersion  which is made before employing a total power detector, i.e., with ``voltage'' signals, shifted to the baseband frequency domain. As a rule, the signals are digitized and digitally processed. The processing performs  phase rotation  opposite to  phase rotation undergone during propagation through  interstellar media. Realization of the filter  can be made in the frequency domain using  FFT or in the time domain using finite-impulse response filter (FIR filter).\\
b) Non-coherent (post-detection) de-dispersion  operates after  total power detection. The total bandwidth of the received signal is divided on many narrow-band sub-bands as in a spectral analyzer and the voltage signals at the outputs of this partial filters are squared. These digitized ``intensity'' signals are time-shifted (with respect to each other) to compensate for the frequency-dependent  delay produced by  the interstellar media.\\
Both de-dispersion methods require some {\it a priori} knowledge of the dispersion measure (DM). This information about DM is usually not available in the search for radio transients. Therefore, when using either method of de-dispersion, several trials of DM must be performed.

2.  Duration matching of transients.\\
  To obtain the maximum possible signal-to-noise ratio which is necessary for reliable detection
 matched filtering must be performed: ``intensity'' signals must be correlated to the template of the expected transient. This is especially important for weak pulses. Usually, neither the form nor even the duration of transient are  known. The wide range of time scales of radio transients and the absence of information about their form   requires the making of multiple trials in order to ``guess'' at least
the duration of a transient. Integration intervals of
``intensity'' signals are tuned to find the optimal  interval   which coincides with the duration of the transient. This is an iterative approximation to the matched filtering.

Both DM trials and iterative matched filtering require considerable computational efforts, especially taking into consideration other search parameters: observational sky frequency and celestial coordinates. In the following sections  an algorithm   alleviating this computational burden will be proposed.

 Essentially the detection of transients is the  detecting  of abrupt changes  in  $\sigma_{s}^{2}$. This means that a permanent monitoring of   $\sigma_{s}^{2}$
  at different time scales should be performed. The algorithm implementing this monitoring should detect  strong and weak changes in amplitude and  short and long changes in time. This task is akin to the   detection  of  change points in stochastic  processes \citep {Basseville}.\\

  Two problems arising in the {\it single-dish single transient} observational situation are considered in this paper:

   1) The amplitudes, time intervals (duration) and moments of arrival of transients are not known. It is therefore difficult  to design a matched filter to detect transients with unknown parameters.  The time scale of transients is very wide.   Very strong transients, like Crab Giant Pulses or Jupiter radio bursts, can be  detected with  simple threshold techniques but weak and rare transients  can be missed if not  all possible parameter values are tried. This deficiency can be  crucial  in the detection of  the non-periodical sporadic unique transients. In radio astronomy   often the signals-of-interest $\sigma_{s}^{2}<<\sigma_{sys}^{2}$ and a considerable amount of raw data samples $n>>1$ are required in order to detect  transients.

    Here  a framework  is proposed for  transients detection algorithms  based on the method of cumulative sums  \citep {Dobben,Basseville} which was first proposed in
  \citep{Page1}.   Let the process under scrutiny  produce observed data $x_{1},x_{2},x_{3}...$. There is a parameter associated with the process. The process is said to be ``in control`` if  the measured mean of the parameter is close to the  {\it target value}. The process, of course, exhibits its own natural variability, for example, in the case of a Gaussian noise. Differences between the observations and the target value will always occur. It is necessary to distinguish between random variations and systematic deviations  due to the process being ``out of control''. This situation often occurs in  quality control in industrial production lines. In our case the parameter-of-interest is the variance $\sigma^{2}$. Page proposed continuous accumulation of  the differences of the measured parameter and the target value - calculation of cumulative sum (cusum). When the running estimate of cusum is below a specified threshold, the process is judged to be ``in control``. As soon as cusum exceeds  the threshold the change point is detected and the ``out of control'' signal is triggered.  Details of the cusum algorithm will be given later. The cusum test uses the combined information of any number of observations, in fact, of all  observations that have been obtained up to the time of testing. The numerical procedure is very simple and is easily mapped on the computer instructions set.

  Cusum is a special type of  sequential probability ratio test (SPRT) developed by \citep{Wald}.  Suppose it is necessary to discriminate between two hypotheses about a parameter-of-interest, the null hypothesis $H_{0}$ being that the process variable is  within tolerable limits and the alternative hypothesis $H_{1}$ being that the process variable is biased by a specified value. At each sampling the ratio of the likelihood of obtaining the observed values under $H_{1}$ and $H_{0}$, respectively, is calculated. If the ratio is large, the alternative hypothesis $H_{1}$ is accepted; if the ratio is small, the null hypothesis $H_{0}$ is accepted; and if the ratio has an intermediate value, the decision is delayed until further observations have produced either a higher or lower ratio of the likelihood. Sequential detection is optimal in the sense that it requires minimal number of observations to trigger a query (average run length), i.e., necessary for the detection of a parameter's change.

  SPRT and its modification - cusum - can be useful tools in the situation of uncertainty about the duration of a transient. Using the above-mentioned terminology, the phenomenon of a transient (the increase and decrease of $\sigma^{2}$) can be described as the ``out of control'' and the  ``in control`` state, respectively.
  Continuous calculation of cusum  of the observational data up to the point of change provides an adaptive behavior and obviates the requirement for duration trials.

  Another problems are RFI.

  2) Man-made radio frequency interferences (RFI) very often produce signals similar to those of natural transients and the algorithms of transients detection must include procedures for distinguishing between  RFI and cosmic signals-of-interest. On a level with the spatial-temporal check (when a transient must be simultaneously registered at the distant antennas pointed to the same radio source in the sky  and taking into account  geometrical delay), a control check at one radio telescope is proposed which allows a distinction to be made  between  the natural transient (Gaussian pdf, pure random noise) and the man-made transients (RFI with non-Gaussian pdf and non-random behaviour).

\section{The detection algorithm}

Raw input data after amplification, filtering and digitization are presented as a sequence of random numbers $x_{i}, i=1..n$, statistically independent and identically
distributed, having normal distribution with zero mean  $p_{k}(x)=\frac{1}{\sigma _{k}\sqrt{2\pi }}\exp (-0.5\sigma _{k}^{2})$ \ $\ k=0,1,$  where $\sigma_{0}$ corresponds to the absence of a transient, $\sigma_{1}$ corresponds to the presence of a transient. The $n$ numbers are stored in the buffer. Starting from   $x_{r}, 1\le r\le n$, pdf $p_{0}$ is changed to the pdf  $p_{1}$. The task is to find this change of  noise variance from $\sigma_{0}$ to  $\sigma_{1}$.

As elsewhere in the detection theory there are two hypotheses: $H_{0}$ - the absence of a change of variance at the interval $1\leq i\leq n$ and $H_{1}$  - the presence of the change point $x_{r}$
 inside the interval $1\le r\le n$. The probability that the volume of data $x$ belongs to the case of $H_{0}$
is equal to $P_{H_{0}}=\prod_{i=1}^{n}p_{0}(x_{i})$, whereas the probability that part of the samples $x_{i}, 1 \leq i \le r$, belongs to $H_{0}$ and the other part of
$x_{i}, r \leq i \leq n$, belongs to $H_{1}$ is equal to $P_{H_{1,r}}=\prod_{i=1}^{r-1}p_{0}(x_{i})\prod_{i=r}^{n}p_{1}(x_{i})$. Therefore, the ratio of
these two probabilities (the {\it likelihood ratio}) is
\begin{equation}
\Lambda _{n}=\frac{P_{H_{1,r}}}{P_{H_{0}}}=\frac{\prod_{i=1}^{r-1}p_{0}(x_{i})\prod_{i=r}^{n}p_{1}(x_{i})}{\prod_{i=1}^{n}p_{0}(x_{i})}=\prod_{i=r}^{n}\frac{p_{1}(x_{i})}{p_{0}(x_{i})}
\end{equation}
This value has its maximum at  $r$ if there is a change of variance in the data.

\subsection{Detection with a known change  moment and duration}

Let us suppose that the change moment  $r$ and the duration of transient $n-r=N$ are known, then  the  {\it matched} filter can be applied for the detection of such a transient. The indexes in sums in this subsection will span from 1 to N  limiting only the length of the transient.
$\Lambda _{N}$ is compared with the threshold  $A$ and if
\begin{equation}
\Lambda _{N}=\prod_{i=1}^{N}\frac{p_{1}(x_{i})}{p_{0}(x_{i})}\geq A,
\label{eq1}
\end{equation}
$H_{1}$ is chosen, i. e., the change in $\sigma$ is detected. The  value of  threshold  $A$ is chosen to minimize two kind of errors: the error of the first type is to give a false alarm, i.e., to make the decision in favor of $H_{1}$ when $H_{0}$ is valid, and the error of the second type  is to miss the change point, i.e., to make the decision in favor of $H_{0}$ when $H_{1}$ is valid. The error probability of the first type is denoted by $\alpha$ and that of the second type is denoted by $\beta$. The value $1-\beta$ is the probability of detection. One of the conventional ways of choosing $\Lambda _{N}$ is to maximize $1-\beta$ for a given $\alpha$ (Neyman-Pearson criterion, \citep{Whalen}).

 Logarithm of (\ref{eq1}) gives:
\begin{eqnarray}
ln(\Lambda _{N})=\ln (\prod_{i=1}^{N}\frac{p_{1}(x_{i})}{p_{0}(x_{i})})=\sum_{i=1}^{N}\ln [p_{1}(x_{i})]-\sum_{i=1}^{N}\ln [p_{0}(x_{i})]=\nonumber\\
\sum_{i=1}^{N}\ln \frac{\sigma _{0}}{\sigma _{1}}+\frac{\sigma _{1}^{2}-\sigma _{0}^{2}}{2\sigma _{0}^{2}\sigma _{1}^{2}}\sum_{i=1}^{N}x_{i}^{2}\geq ln(A),
\end{eqnarray}
or:
\begin{eqnarray}
\frac{1}{N}\sum_{1}^{N}x_{i}^{2}\geq \frac{\frac{1}{N}\ln (A)-\ln \frac{\sigma _{0}}{\sigma _{1}}}{\frac{\sigma _{1}^{2}-\sigma _{0}^{2}}{2\sigma _{0}^{2}\sigma _{1}^{2}}}
\label{eq2}
\end{eqnarray}
The left part of (\ref{eq2}) corresponds to the usual  radiometric output (the averaged sum of the sample's squares), or the {\it energy} detector when the moment and duration of the transient are known. The threshold in the right part of (\ref{eq2}) depends on $\sigma _{0}$  and $\sigma _{1}$, but for the signal increment $\Delta \sigma ^{2}=\sigma _{1}^{2}-\sigma _{0}^{2}<<\sigma _{0}^{2}$ and large $N>>1$ which is the typical case in radio astronomy, we can compare $\frac{1}{N}\sum_{1}^{N}x_{i}^{2}$
with the value $h=\sigma _{0}^{2}+k_{0}\sigma _{0}^{2}\sqrt{2/N}$, where the coefficient $k_{0}$ is chosen to satisfy the probability of false alarms $\alpha$ to be equal to the prescribed  value. For example, for $\alpha=0.05,k_{0}=1.645$,   for $\alpha=0.01,k_{0}=2.33$,   for $\alpha=0.001,k_{0}=3.09$.\\
The probability of detection for the given threshold $h$ and the signal+system noise variance  $\sigma _{1}^{2}$ is
\begin{equation}
P_{\det }(h,\sigma _{1})=0.5-0.5erf(\frac{h-\sigma _{1}^{2}}{2\sigma _{0}^{2}\sqrt{1/N}}),
\end{equation}
where $erf(x)=\frac{2}{\sqrt{\pi }}\int_{0}^{x}\exp (-t^{2})dt$.

\begin{figure}
\includegraphics[width=80mm, height=60mm]{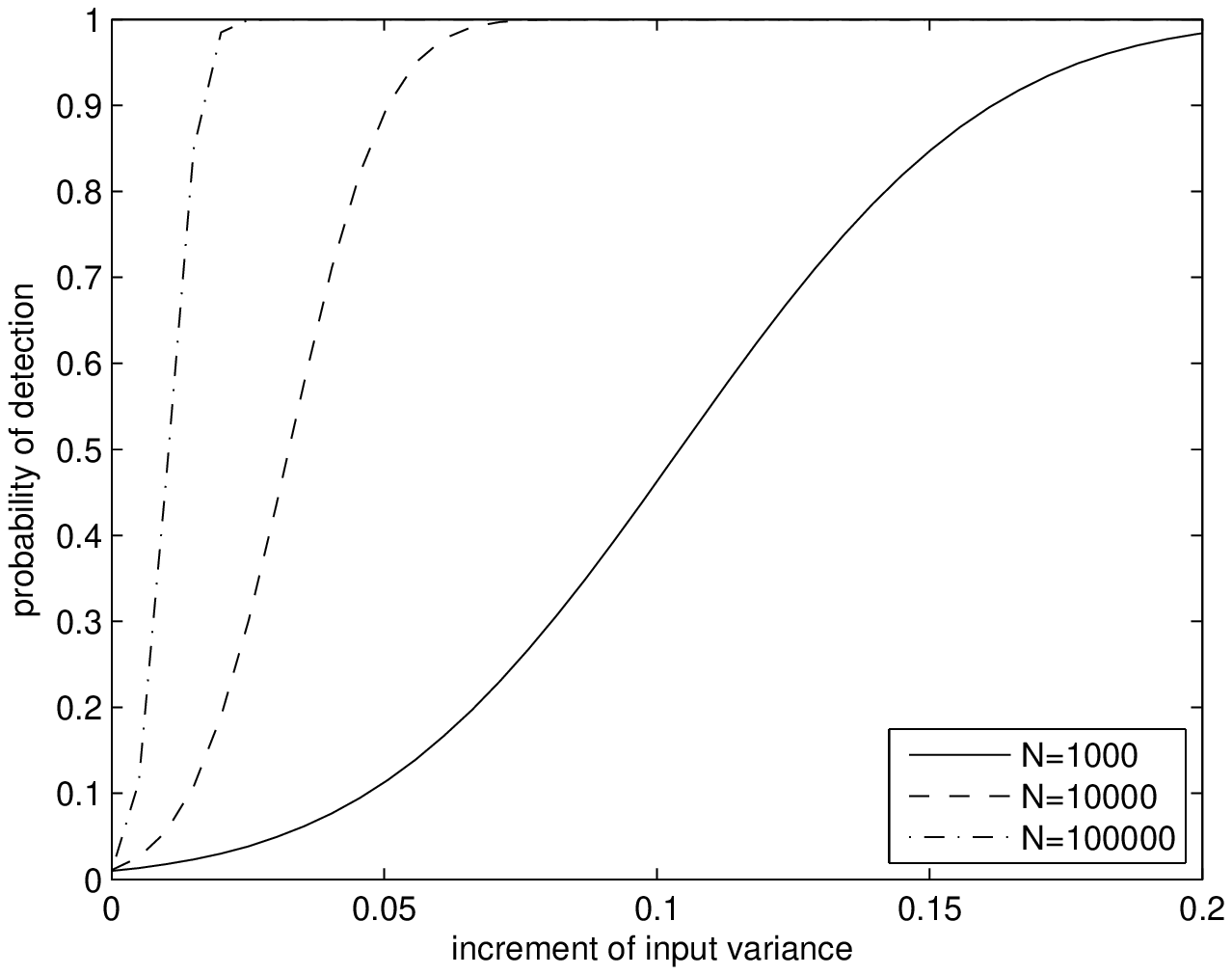}
\includegraphics[width=80mm, height=60mm]{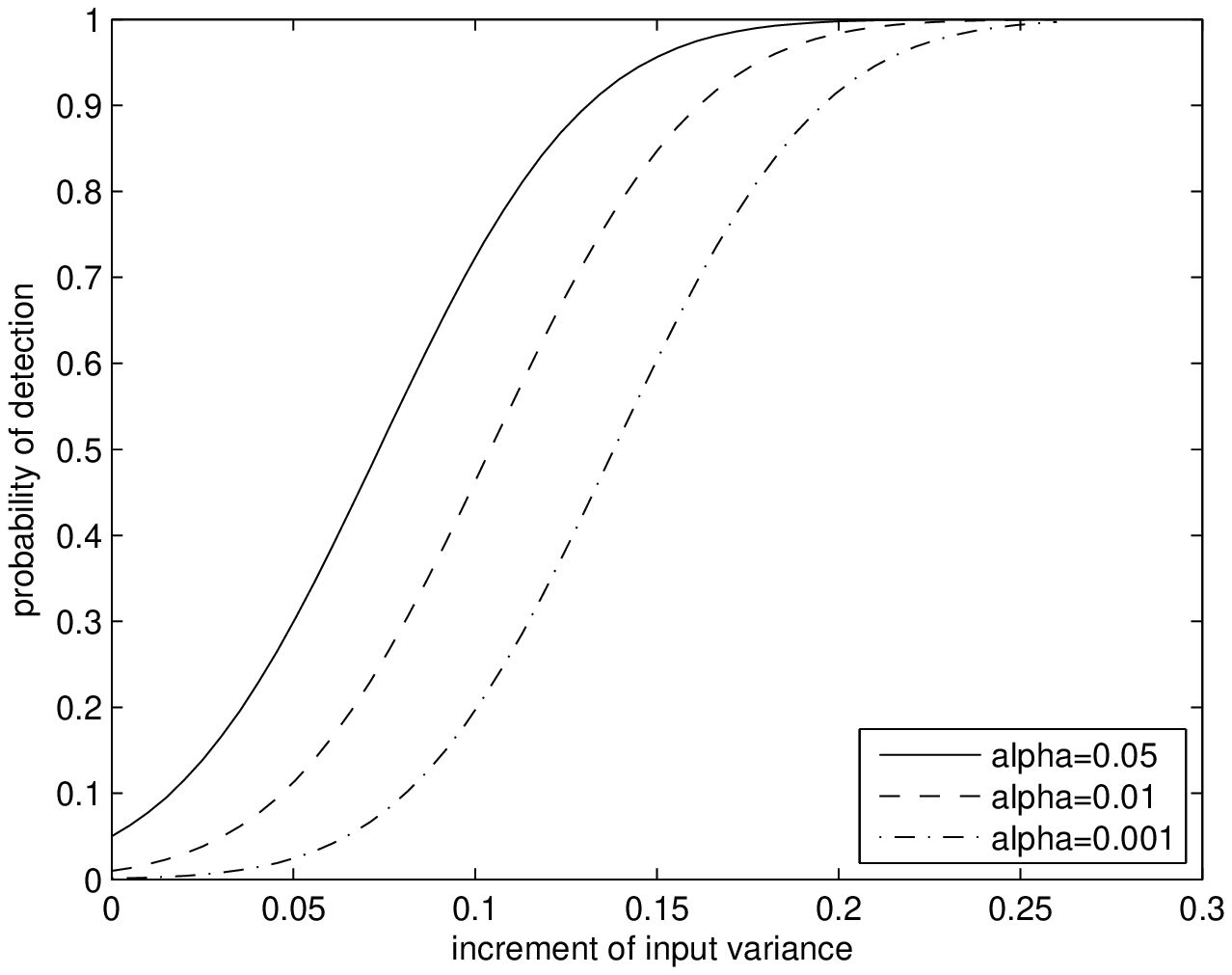}
\includegraphics[width=80mm, height=60mm]{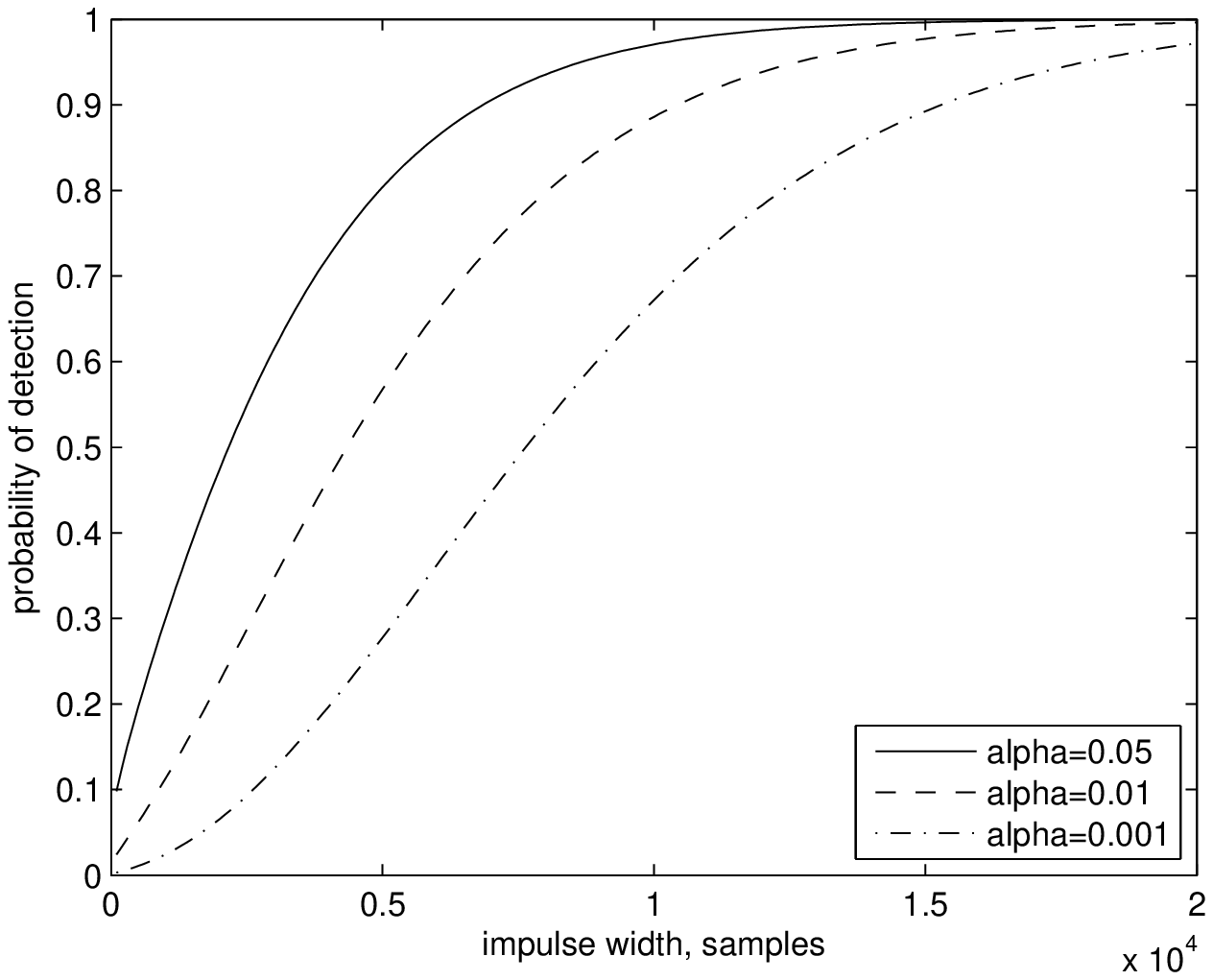}
\caption{Probabilities of detection of  noise-like transient with normal pdf and known duration and time of arrivel, the upper figure: probability of false alarms $\alpha=0.01$, and the duration (the number of transients samples $N$) is the parameter, $N=10^{3},N=10^{4}$ and $N=10^{5}$, horizontal axis is the transient amplitude $\Delta \sigma ^{2}=\sigma _{1}^{2}-\sigma _{0}^{2}$; the middle figure:  probabilities of detection when $\alpha$ is the parameter and $N=10^{3}$; the lower figure:  probabilities of detection as  functions of the number of samples $N$ and $\sigma_{1}=1.0245$ and $\alpha$ is the parameter.}
\label{detN}
\end{figure}

Fig. \ref{detN}  shows the probabilities of  detecting  transients: the top figure corresponds to $\alpha=0.01$  and the duration (the number of transients samples $N$) is the parameter, $N=10^{3},N=10^{4}$ and $N=10^{5}$, the horizontal axis is the transient amplitude $\Delta \sigma ^{2}=\sigma _{1}^{2}-\sigma _{0}^{2}$; the middle figure shows the  probabilities of detection $1-\beta$ when $\alpha$ is the parameter and $N=10^{3}$; the lower figure demonstrates the dependence of $1-\beta$  of the transient duration $N$ when $\sigma_{1}=1.0245$ and $\alpha$ is the parameter.

All these curves   show the best possible outcome of the testing of the hypothesis when the moment of arrival of the transient and its duration are known.
In real life this is not the case, and a bank of matched filters is necessary, each tuned to the particular $r$ and $N$ in the range of expected values. These trials on $r$ and $N$ correspond to the generalized likelihood ratio test which are  additional computational burdens on the inevitable trials on  the dispersion measure.

\begin{figure}
\includegraphics[width=80mm, height=60mm]{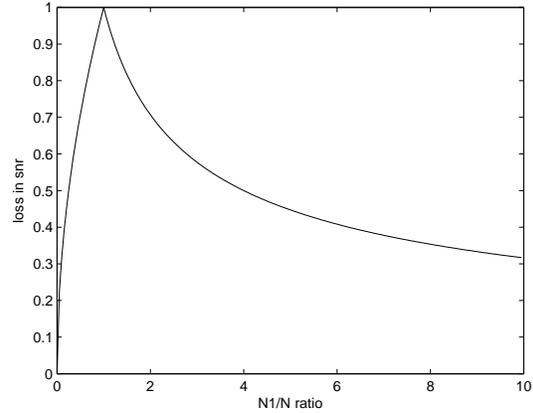}
 \caption{The relative snr due to the mismatch  in the number of samples $N_{1}$ detecting the transient  which duration interval corresponds to $N$ samples.}
 \label{wrong_n}
\end{figure}

 Choosing the wrong $N$ may significantly  worsen the probability of detection. Let us consider the situation when the ``rectangular'' signal with the amplitude $a$ and duration number $N$ of samples must be detected at the background of noise with {\it rms}=$\sigma$. For the known $N$, the signal-to-noise ratio after the averaging of
 $N$ samples of the mixture ``signal+noise'' is $snr_{N}=\frac{a\sqrt{N}}{\sigma }$. \\
 For $N_{1}>N$ the averaged signal is $a\frac{N}{N_{1}}$ and the rms of the noise is $\sigma \frac{1}{\sqrt{N_{1}}}$. Therefore, the signal-to-noise ratio is $snr_{N_{1}}=\frac{aN}{\sigma \sqrt{N_{1}}}$. \\
 For $N_{1}<N$  the signal-to-noise ratio is $snr_{N_{1}}=\frac{a\sqrt{N_{1}}}{\sigma }$. The ratio $\frac{snr_{N_{1}}}{snr_{N}}$ shown in Fig.  \ref{wrong_n}  as the function of $k_{N_{1}}=N_{1}/N$ gives an impression about  losses due to the mismatch in the duration of the impulse. The decrease in the number of detectable radio sources is proportional to $k_{N_{1}}^{3/4}$ for $k_{N_{1}}<1$ and to $k_{N_{1}}^{-3/4}$ for $k_{N_{1}}>1$.

In order to deal with  the more realistic situation of unknown $r$ and $N$, we will now consider another approach.

\subsection{Sequential analysis}

Until now the test with a fixed sample size  $N$ (transient duration) has been considered.  The sequential probability ratio test  does not require the sample size to be fixed {\it a priori}  but instead uses the data available at each particular moment, \citep{Wald}. The main idea of SPRT is to compare the likelihood ratio calculated for each $i=1..n$ with two thresholds:\\
accept $H_{0}$ if $\Lambda _{n}\leq B$,\\
accept $H_{1}$ if $\Lambda _{n}\geq A$,\\
continue to observe if $B \leq \Lambda _{n}\leq A$.

\begin{figure}
\includegraphics[width=84mm, height=65mm]{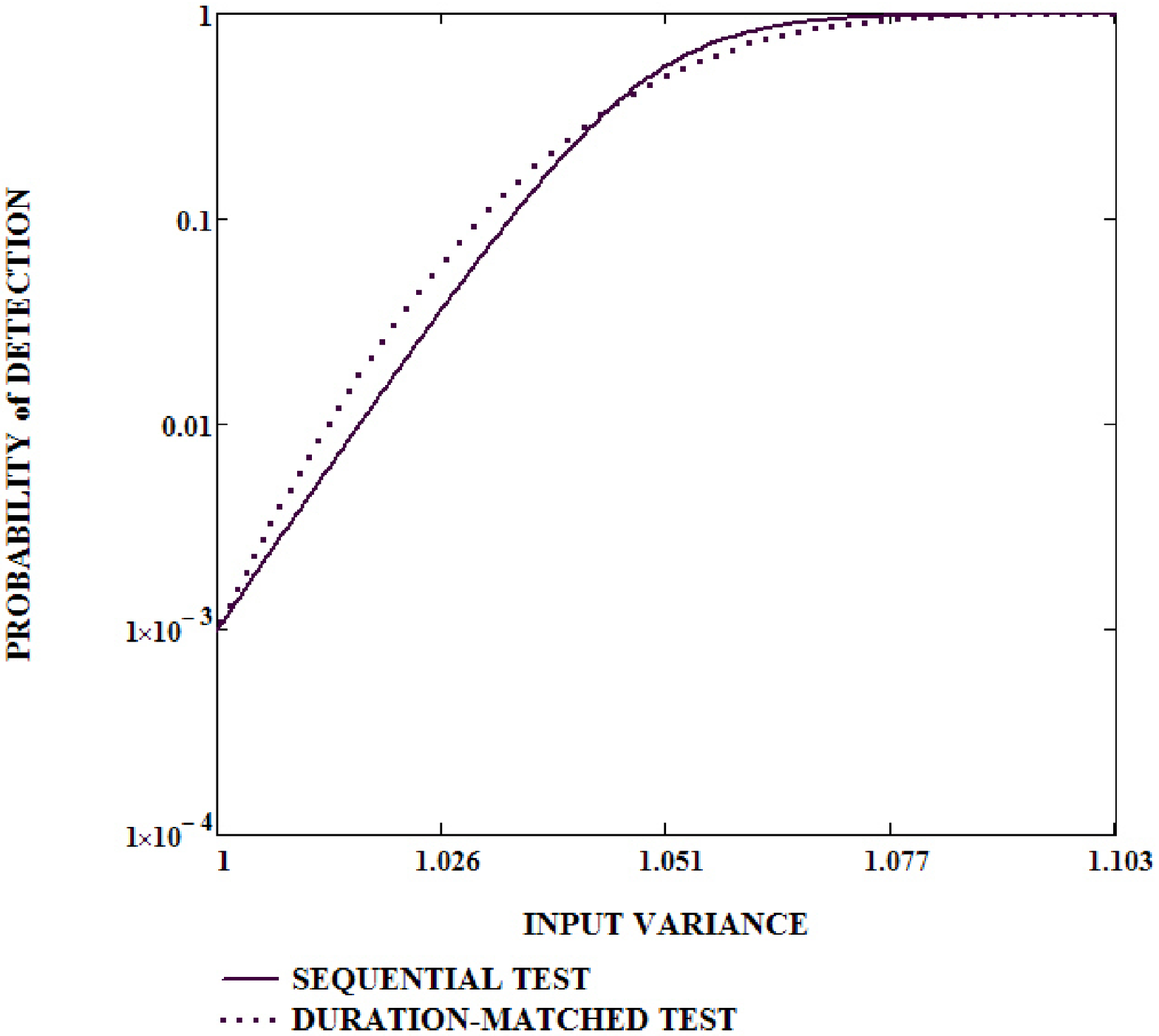}
\includegraphics[width=84mm, height=65mm]{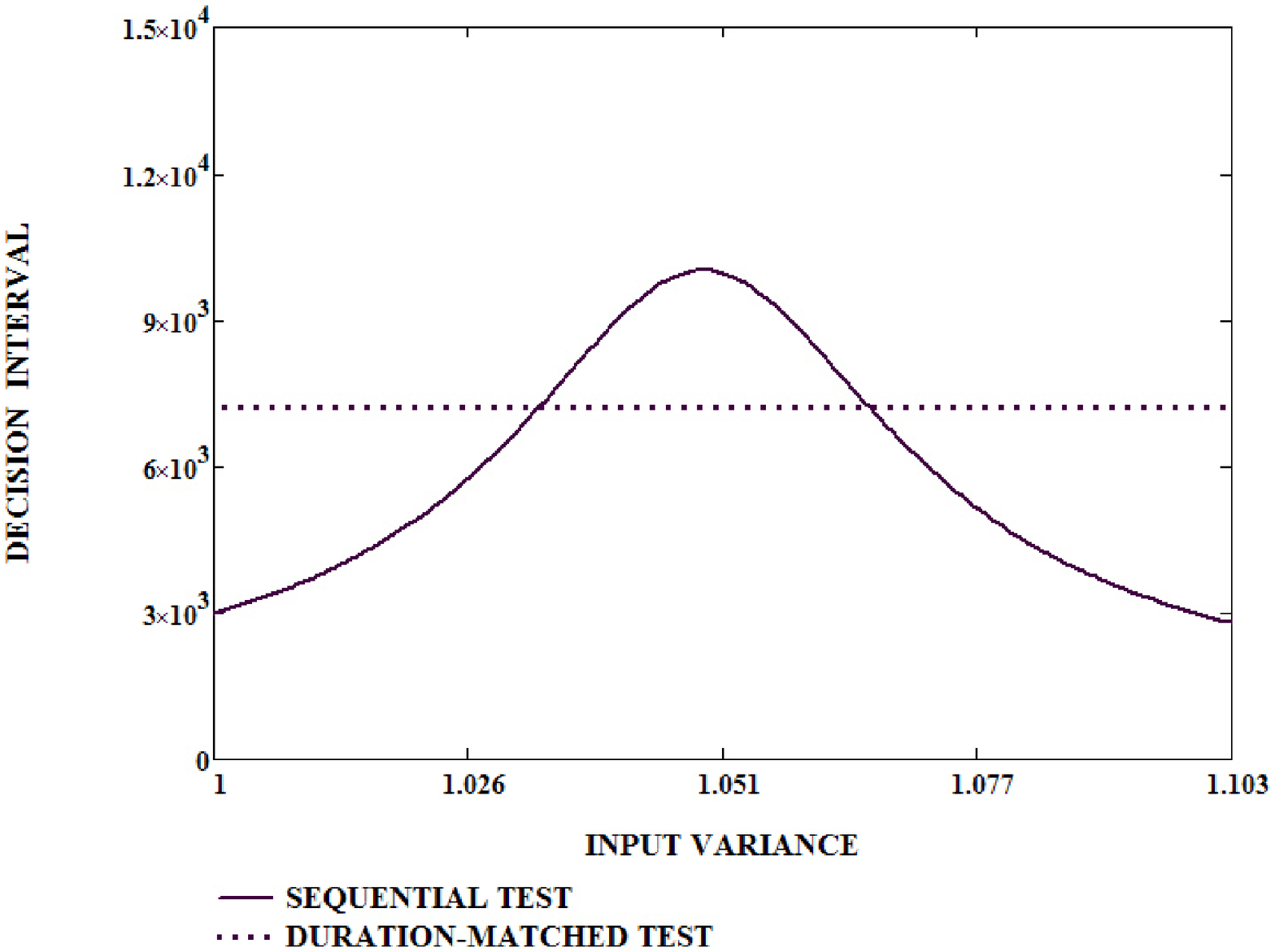}
\includegraphics[width=84mm, height=65mm]{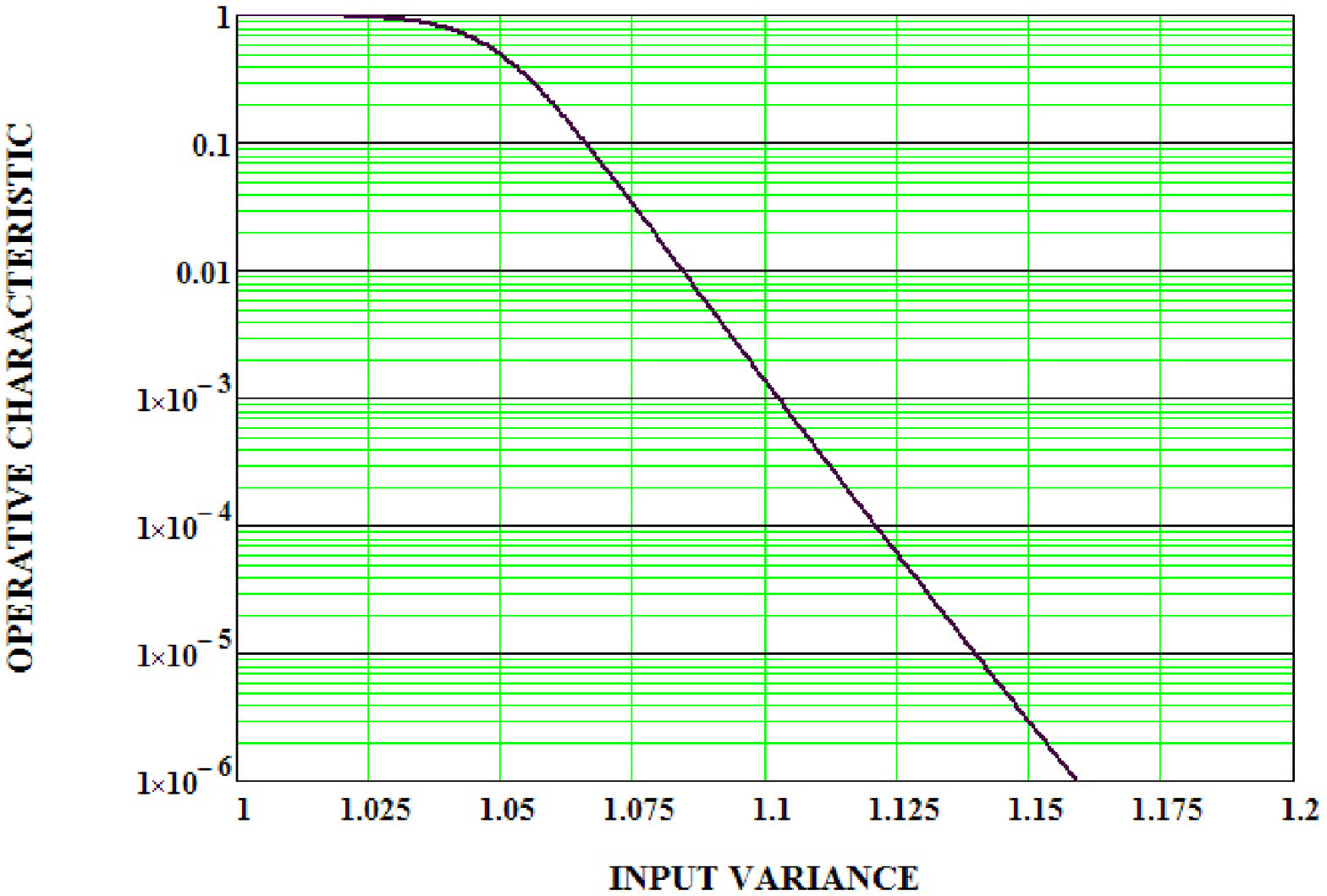}
\caption{The upper panel: probability of detection of the noise-like signal for the sequential test (solid line) and duration-matched  test (dotted line) as  functions of  input variance $(\sigma_{1})^{2}, \sigma_{0}=1.0$,  $\alpha=10^{-3}$, the number of the samples of the signal $N=7200$; the middle panel: average decision time for the sequential test (solid line) as  function of  input variance,the dotted line - number of signal samples;
the lower panel: operative characteristic of SPRT, $\sigma_{1}^{2}=1.103, \alpha=10^{-3}, \beta=10^{-3}$.}
\label{wald}
\end{figure}

The thresholds $A$ and $B$ are chosen as
\begin{equation}
 A=\frac{1-\beta }{\alpha }, B=\frac{\beta}{1-\alpha}.
 \end{equation}
  where  $\alpha$  and $\beta$ are the error probabilities of the first  and second type, respectively, see section 2.1.\\
  Fig. \ref{wald} (the upper panel)  shows the probabilities of detection and average decision time for SPRT tuned for $\alpha=10^{-3}$ and $\beta=10^{-3}$ and for  matched detection: the upper panel - the probability of detection of a noise-like signal for the sequential test (solid line) and the duration-matched  test (dotted line) as  functions of  input variance $(\sigma_{1})^{2}, \sigma_{0}=1.0$,  $\alpha=10^{-3}$, the number of the samples of the signal $N=7200$. There is no big difference between these two curves, whereas
  the middle panel  shows the average decision time for the sequential test (solid line) as the function of the input variance, the dotted line - number of signal samples $(N=7200)$. \\
  There are three regions in the middle panel of Fig. \ref{wald}: weak signals (left side), strong signals (right side) and intermediate signals. The average   number of samples  which is necessary to make the decision is less than that of
   the matched detector for  weak and strong signals and higher in the intermediate case. In the area of interest where  the probability of detection is close to 1 the gain in the decision time  for SPRT is obvious. Therefore, the number of observations   before a decision  is variable and depends on the observational situation. The mean of this random value is the {\it average sample number}.

  The {\it operative characteristic} introduced by Wald for SPRT is  a useful performance parameter. In our case it is the probability of accepting hypothesis $H_{0}$ as a function of $(\sigma_{1})^{2}$:  $L(\sigma_{1}^{2})$. When $\sigma_{1}^{2}=\sigma_{0}^{2}$, i. e., in the absence of signal, $1-L(\sigma_{0}^{2})=\alpha$ (probability of false alarms).
In the presence of signal, $\sigma_{1}^{2}>\sigma_{0}^{2}, L(\sigma_{1}^{2})=\beta$ (probability of missing the signal) and $1-L(\sigma_{1}^{2})=P_{det}$ (probability of signal detection).  It is visible from the  lower panel of Fig. \ref{wald} that the detector
provides $\alpha$ for $\sigma_{1}^{2}=1.0$ and $\beta$ for $\sigma_{1}^{2}=1.103$ both equal to $10^{-3}$ as was planned and the quality of detection
 improves with the growth of the signal. The average decision time in the middle panel also reduces with the signal's amplitude. Therefore,
if  SPRT is tuned for $\sigma_{1}^{2}$ which is chosen for the minimal expected signal, no deviation   $\sigma^{2}>\sigma_{1}^{2}$ is  detrimental.

\subsection{Cumulative sum method}
\begin{figure}
 \includegraphics[width=84mm, height=50mm]{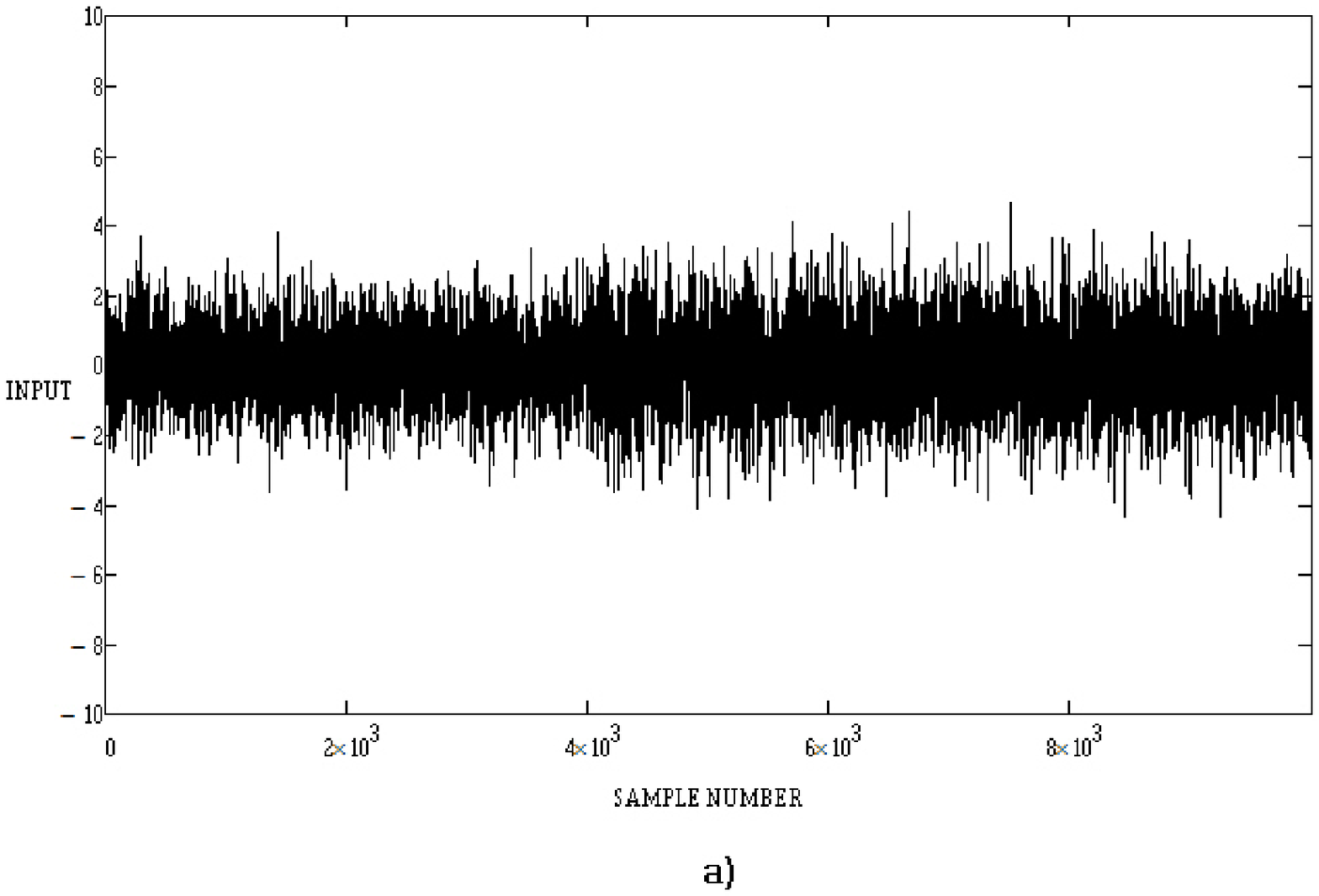}
 \includegraphics[width=84mm, height=50mm]{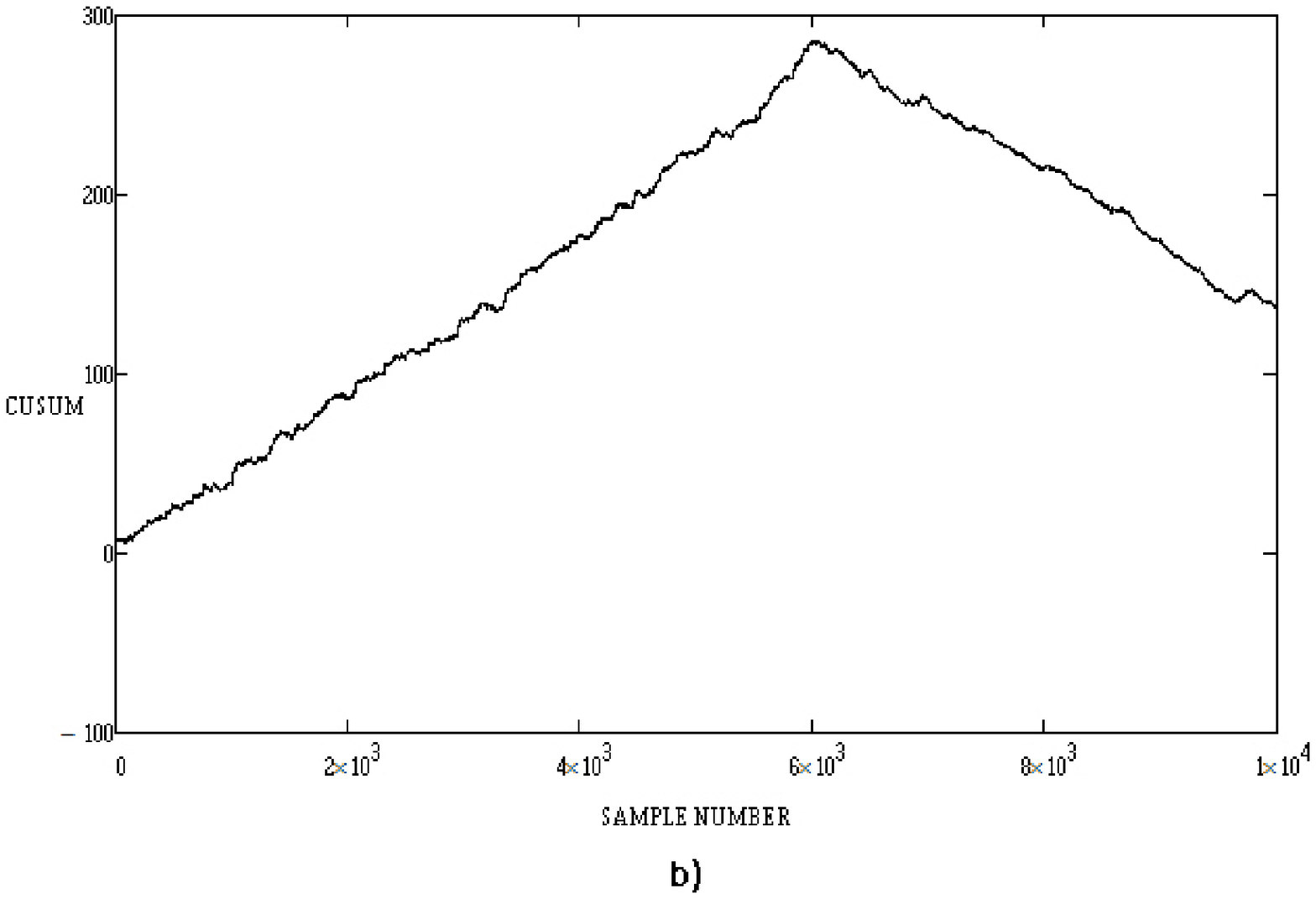}
 \includegraphics[width=84mm, height=50mm]{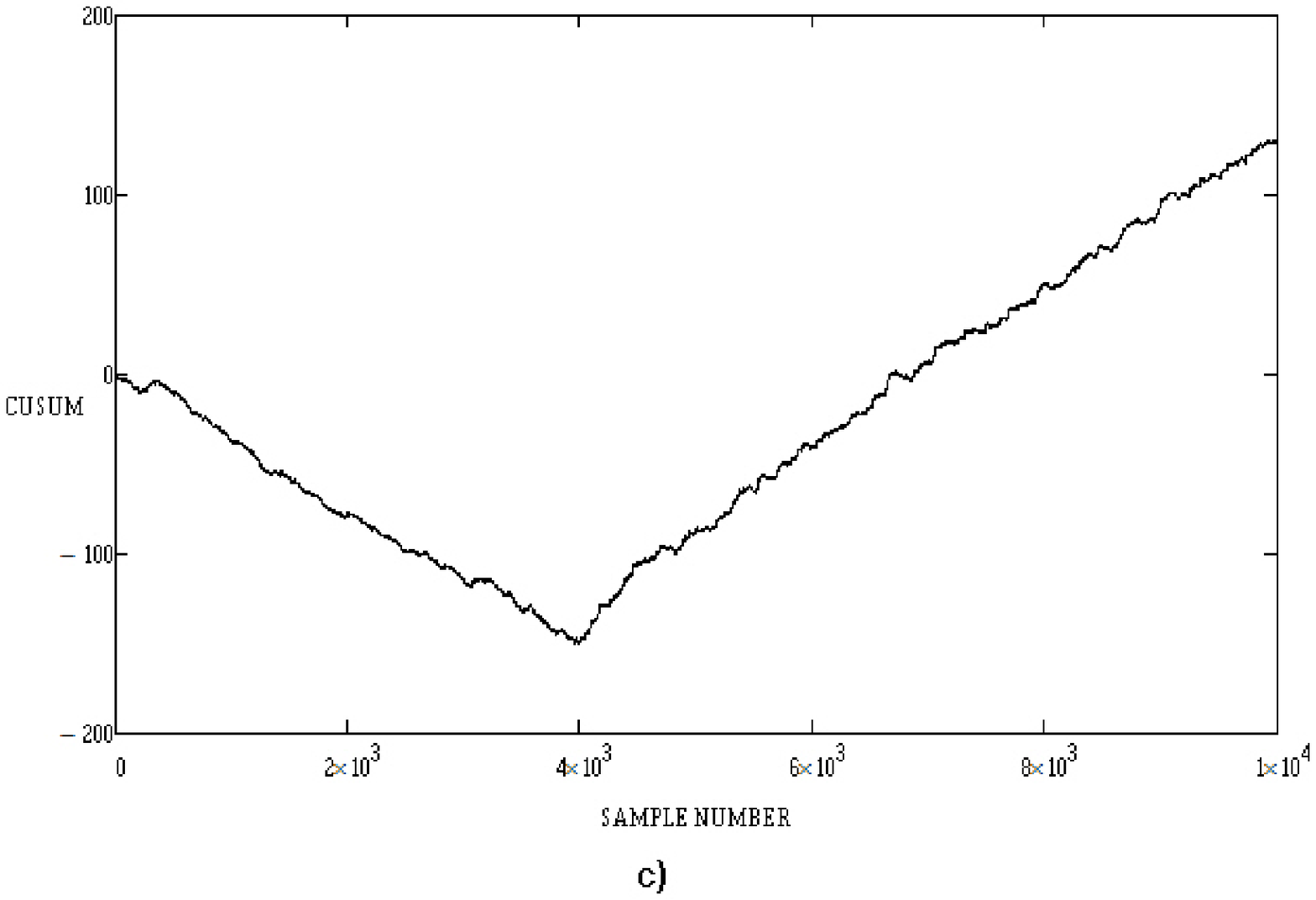}
 \includegraphics[width=84mm, height=50mm]{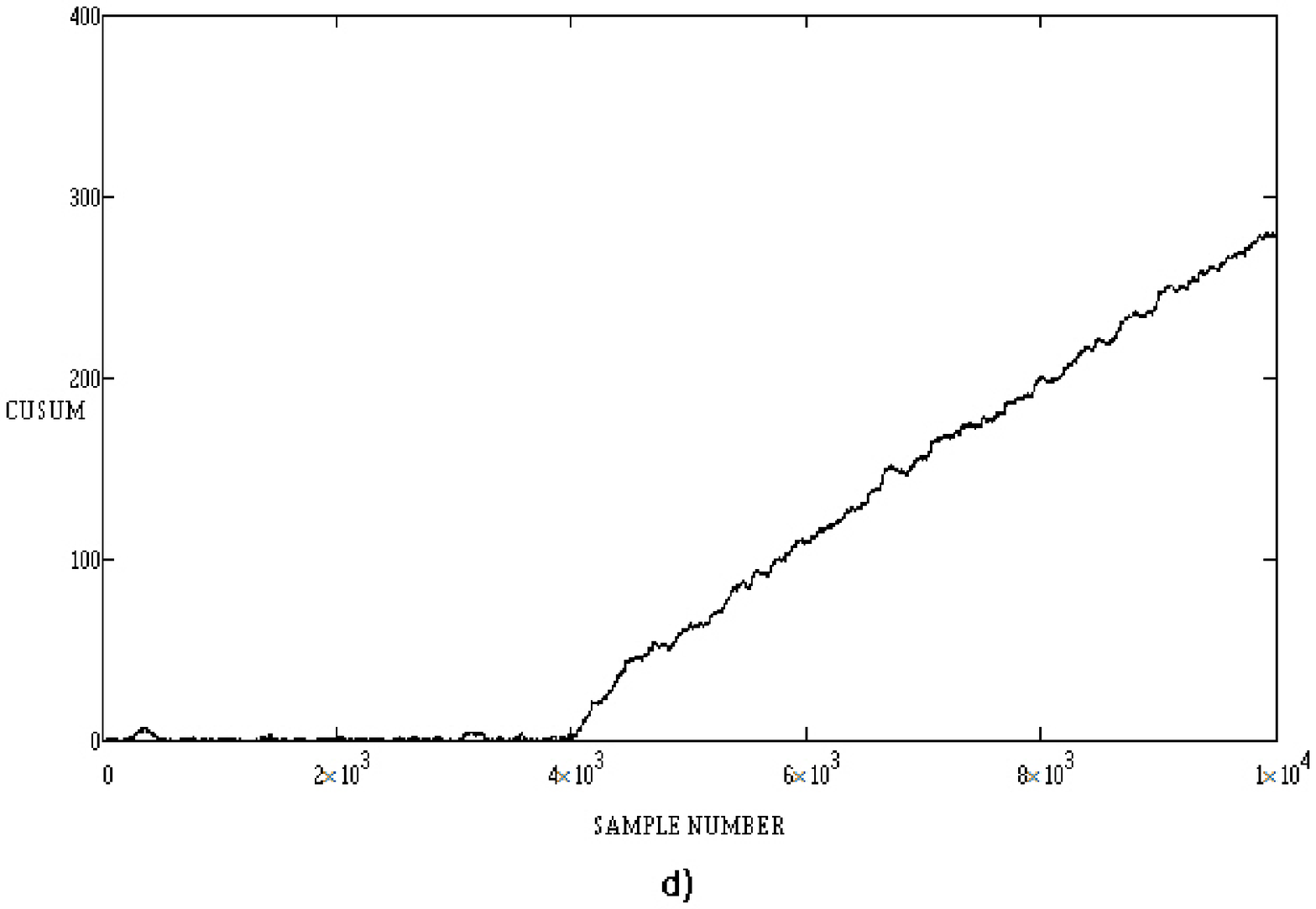}
 \caption{a) input data, $n=10^{4}$,  variance before change $\sigma_{0}^{2}=1.0$, variance after change at point $r=4000, \sigma_{1}^{2}=1.49$;
 b) backward cumulative sum $S_{r}^{n}$; c) forward cumulative sum $S_{1}^{r}$; d) modified forward cusum.
}
\label{input}
  \end{figure}

The next improvement of the detection algorithm is the following modification of SPRT.  Let us return to the expression (3) and denote
\begin{equation}
k=\frac{2\ln (\frac{\sigma _{0}}{\sigma _{1}})\sigma _{0}^{2}\sigma _{1}^{2}}{\sigma _{0}^{2}-\sigma _{1}^{2}}.
\label{k}
\end{equation}
The sum in (3) is rewritten in the following sequence of recursive {\it cumulative} sums:
\begin{eqnarray}
S_{0}^{r}=0\nonumber\\
S_{i}^{r}=S_{i-1}^{r}+x_{i}^{2}-k, i=1..n
\end{eqnarray}
The behavior of  $S_{i}^{r}$ is shown in Fig. \ref{input}.

Fig. \ref{input}a shows  $n=10^4$ samples of Gaussian noise. Samples numbered under  $r=4000$  have  variance $\sigma_{0}^{2}=1.0$ and samples numbered above $r=4000$  have variance $\sigma_{1}^{2}=1.49$.
 Fig. \ref{input}b shows the cumulative sum (cusum) $S_{i}^{r},i=n..1$, i.e., the calculation of cusum goes from the end of the data block to the beginning.
The change point sample $\widehat{r}$ can be found as
\begin{equation}
\widehat{r_{bwd}}=\arg \max_{1\leq i\leq n}(S_{i}^{r}).
\end{equation}
The order of summation can be reversed: from the beginning to the end, i.e., to calculate cusum $S_{i}^{r}, i=1..n$. In this case cusum looks as in Fig. \ref{input}c and
the change point sample $\widehat{r}$ can be found as

\begin{equation}
\widehat{r_{fwd}}=\arg \min_{1\leq i\leq n}(S_{i}^{r}).
\end{equation}

The change point detection rule is the comparison of  $S_{i}^{r}, i=1..n$ with the two thresholds $h_{0}$ and $h_{1}$: when $S_{i}^{r}<h_{0}$ the hypothesis $H_{0}$ is accepted and when $S_{i}^{n}>h_{1}$ the hypothesis  $H_{1}$ is accepted. The intermediate  values of  $S_{1}^{r}$ dictate the continuation of testing (cusum calculation). This is the philosophy of  sequential analysis \citep{Wald}. \\
Page proposed  restarting the algorithm as long as the previously taken decision is  $H_{0}$ and also proposed  making the lower threshold $h_{0}=0$,
\citep{Page1}.
It was shown later
\citep{Shiryaev, Lorden}
 that this is the optimal value for $h_{0}$. Taking this into consideration, the  change point detection rule can be modified in the following manner:
\begin{equation}
\widehat{r}=\min\lbrace r:(S_{1}^{r})^{+}\geq h_{1}\rbrace,
\label{cs1}
\end{equation}
where
\begin{equation}
(S_{1}^{r})^{+}=\max(0,S_{1}^{r}).
\label{cs2}
\end{equation}

Fig. \ref{input}d shows this modified cusum with the visible change point at $i=4000$. The large amplitude of change has been chosen deliberately to better illustrate  the behavior of cusum .

  If a change does not exactly correspond to $\sigma_{1}$, the algorithm is not optimal.  In radio astronomy practice, the value of $\sigma_{1}$ in (\ref{k}) can be chosen as a minimum expected change of variance which  basically can be estimated from the knowledge of the system parameters: the antenna's effective area, the system temperature. For  values larger than $\sigma_{1}$ the detection procedure is not optimal but  is acceptable, because the growth of the change value yields a reduction in ARL and the advantage of an optimal procedure is not great.

The detection performance of cusum is characterized by the probability of false alarm. The second kind of error $\beta$ is not considered in the cusum analysis because  hypothesis $H_{0}$ is never accepted: each time the cusum is equal to zero, the algorithm restarts. The cusum test will eventually reject the hypothesis $H_{0}$. The number of  samples up to this rejection of $H_{0}$ is also a random variable, as in SPRT, and  is called the {\it run length}. Similarly to SPRT, the following  parameter is introduced for cusum:
the average number of samples from the starting point up to the point at which the decision threshold $h=h_{1}$ is crossed  is called the {\it average run length}, ARL. The ARL, the average sample number $\widehat{n}$ and the operative characteristic $L $ are related by
\begin{equation}
ARL(\sigma _{0},\sigma _{1})=\frac{\widehat{n}(\sigma _{0},\sigma _{1})}{1-L(\sigma _{0},\sigma _{1})}
\end{equation}
 \begin{figure}
 \includegraphics[width=84mm, height=60mm]{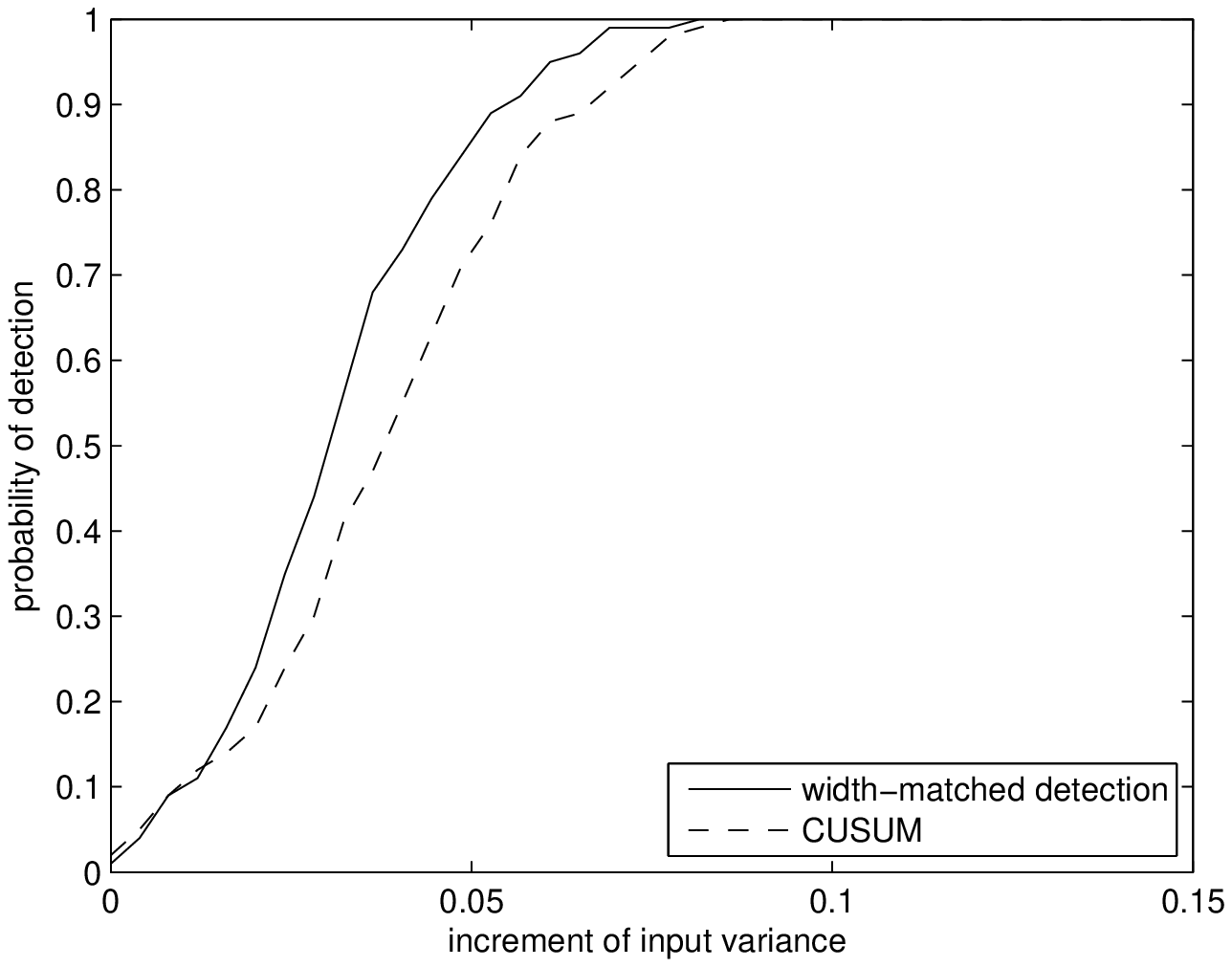}
 \includegraphics[width=84mm, height=60mm]{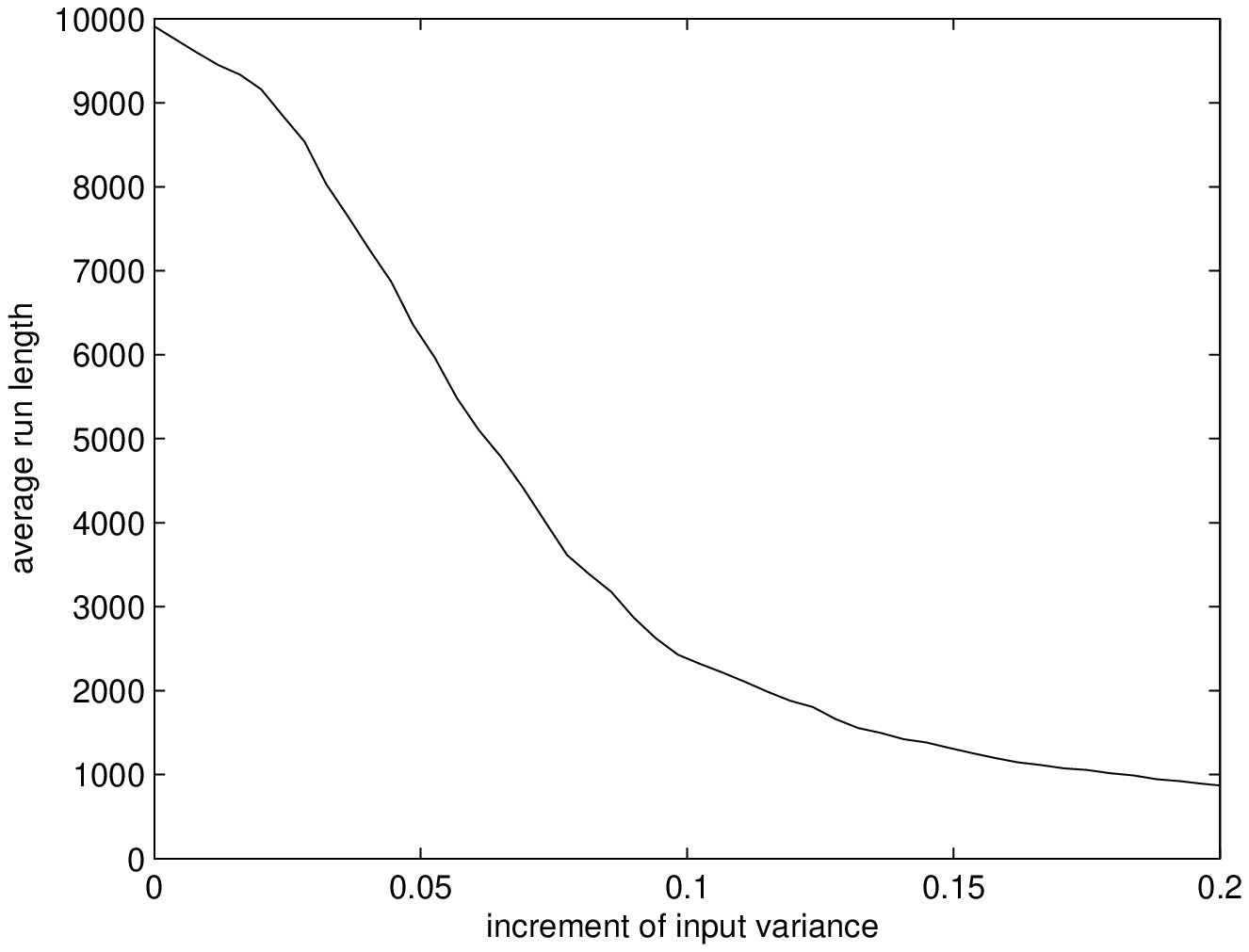}
\caption{Upper panel: probability of detection for cusum and matched detector  for the number of samples $N=10^{4},\alpha=0.01$; lower panel: the average run length for cusum.}
\label{cusum}
  \end{figure}
A theoretical calculation of ARL is a complicated  procedure. Details of these calculations and tables with  useful practical results can be found in \citep{Dobben}, \citep{Basseville} and \citep{Hawkins}. Here we are interested in comparing  cusum performance with the matched detector and the computer simulation was performed for the purpose of demonstration.

  \subsection{Computer simulation}

  The following computer simulations were made to compare cusum with  width-matched detection. The number of samples of  noise with  normal pdf  ${\cal N} (0,\sigma)$ is $N=10^{4}$. In the absence  of signal $\sigma=1.0$ for all $N$ samples. A change of $\sigma>1.0$ must be detected. The width-matched algorithm integrates all $N$ samples and the decision about the presence of a signal (any change in $\sigma>1.0$) is made when the estimate of  variance exceeds the threshold $h=\sigma _{0}^{2}+k_{0}\sigma _{0}^{2}\sqrt{2/N}$, where $\sigma _{0}=1.0$ and $k_{0}=3.09$ is chosen to keep the probablity of false alarm at $\alpha=10^{-3}$.

  The cusum algorithm  (\ref{cs1}) and (\ref{cs2}) is tuned with $\sigma _{0}=1.0$ and $\sigma _{1}=1.05$.  The threshold $h_{1}$ was also chosen to keep
  $\alpha=10^{-3}$. Both tests (matched detection and cusum detection) were  repeated 100 times. \\
  Averaged results are given in Fig. \ref{cusum}.
  The upper panel shows that the probabilities of detection as  functions of the signal increment 
  for cusum are slightly lower in the region of uncertain detection and are close to $1.0$ at the same value as for the width-matched detector. \\
  The lower panel shows the dependence of the average run length on the signal power. The number of samples which are necessary  for  signal detection decreases and at the point of $P_{det}\approx1.0$
  is four times less than $N=10^{4}$. ARL continues to decrease further with the  growth of the signal, while $N$ remains constant being preset for the matched detector. This is the important  property of cusum.\\
  It must also be mentioned that  the comparison of different types of transient detectors made in \citep{wang}  showed the advantage   of using cusum.

\subsection{Examples from observations}

The pulsar machine PUMA-2 installed at WSRT allows  raw data to be stored in the 20MHz bandwidth. The radio telescope works in tied-array mode in which all 14 signals from antennas are added in phase, i.e., there is one output as for a single dish. The 20MHz baseband signals are digitized (8bit, $4\cdot10^{7}$ samples/sec) and stored in the mass storage system which has  sufficient capacity to support 24 h of continuous observations. Signal processing can  therefore
be undertaken {\it off-line}.

\begin{figure}
 \includegraphics[width=84mm, height=60mm]{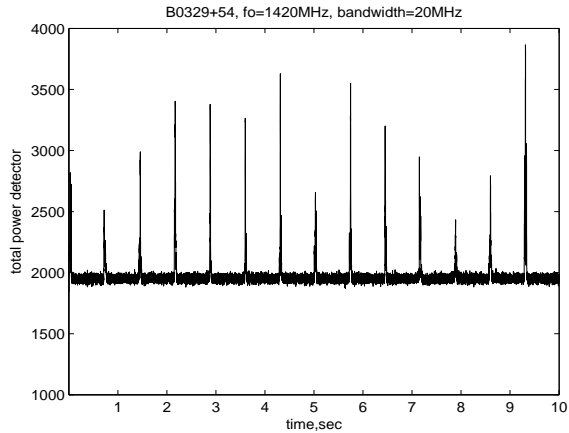}
 \caption{Impulses of pulsar B0329+54 observed at WSRT with PUMA-2. Sky frequency=1420MHz, bandwidth=20MHz, integration time=$5\cdot10^{-4}sec$.}
 \label{puma10}
 \end{figure}

\begin{figure}
 \includegraphics[width=84mm, height=60mm]{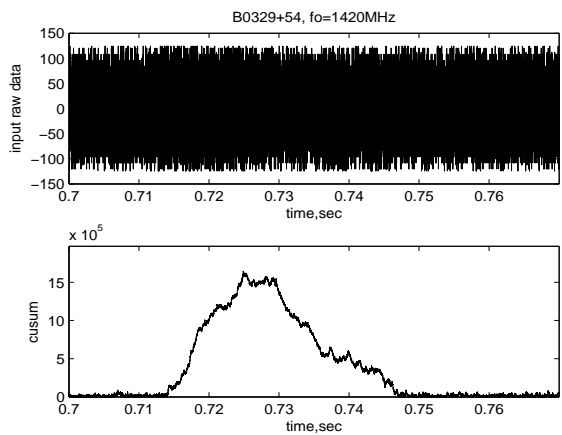}
 \caption{The upper panel:  raw data corresponding to the area of the second impulse in Fig. \ref{puma10}. The lower panel: cusum calculated with the samples of raw data shown in the upper panel.}
 \label{pumaraw}
 \end{figure}

\begin{figure}
 \includegraphics[width=84mm, height=60mm]{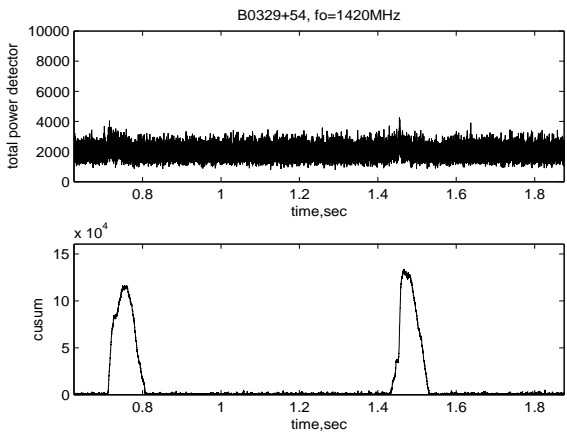}
 \caption{The upper panel:   TPD data corresponding to the area around the second and  third impulses in Fig. \ref{puma10}, integration time=$1.25\cdot10^{-6}sec$. The lower panel: cusum calculated with  samples of the  data shown in the upper panel.}
 \label{puma2}
 \end{figure}

A block of data corresponding to $\approx 10sec$ observation of pulsar B0329+54 at the sky frequency 1420 MHz is used here for  demonstrating  transient detection. Fig. \ref{puma10}  shows the sequence of pulsar impulses: total power detector and   integration at $5\cdot10^{-4}sec$ (20000 samples). The raw data corresponding to the time interval  from 0.7 sec till 0.77 sec (an area around the second impulse in Fig. \ref{puma10})
 is given by the total number of samples $2.8\cdot10^{6}$ which is too large to be represented in one figure. Therefore,  only one sample from each 50
is shown in Fig.  \ref{pumaraw},  (upper panel), i.e., there is   decimation  equivalent to a 50 times reduction of the effective bandwidth: to  0.4MHz.\\
The cumulative sum calculated with these decimated samples is shown in Fig. \ref{pumaraw},  (lower panel) and clearly indicates the pulsar impulse, {\it with no assumption made about its duration}.\\
Fig.  \ref{puma2}  shows another way of using cusum. The output of the total power detector (TPD) with integration over 50 samples  is shown in the upper panel of Fig. \ref{puma2}. This waveform corresponds to the second and third impulses in Fig. \ref{puma10}  (from 0.7sec till 1.9sec). Insufficient integration  reveals some small increase in  noise variance at the positions of impulses. The lower panel shows the cumulative sum calculated using the data represented in the upper panel of Fig. \ref{puma2}. And again, no assumption was made about the duration of impulses. The  ability to detect  these two transients is clear.\\
Looking at the lower panel of Fig. \ref{puma2} one could imagine that this waveform could equally well be obtained with a smoothing algorithm or with a low-pass digital filtering procedure. But the principal advantage of cusum is the  freedom from any   choice of  matching integration parameters: the smoothing interval or the cut-off frequency of the digital filter.

Until now the one-dimensional  approach was  considered. The usual practice includes the spectral analysis of the data with FFT or filter banks. It means that there is a two-dimensional array: a time-frequency  plane. This multi-narrow-band filtering is useful for  flagging of RFI and de-dispersion. In principle, cusum can be calculated in each frequency channel, if an expected  transient is strong enough. Otherwise,  the channels free from RFI after  de-dispersion can be averaged and analyzed with cusum algorithm. \\

\subsection{Transient detection in the presence of dispersion}

  Transients are  broadened by propagation effects and their amplitude decreases. To compensate this effect  trials of  different values of dispersion measures  are usually made in order to  concentrate all possible pulse energy into one narrow burst. After each DM trial  the time integration (duration matching)  is performed to detect the transient. Cusum can be used at this stage to reduce the number of trials of transient's durations.

In the ideal case of  phase-only distortions during propagation, the total energies of the broadened pulse and the initial narrow pulse (without dispersion) are equal.  When the duration $W_{i}$ of a pulse is known  and its amplitude is equal to $AMP_{i}$ the probability of detection can be estimated by the well-known formula, \cite{Whalen}:
\begin{equation}
P_{\det }=1-\frac{1}{\sqrt{2\pi }}\int_{\sqrt{E/N_{0}}}^{\infty }\exp (-u^{2}/2)du,
\end{equation}
where $E=AMP^{2}_{i}W_{i}$ is the pulse energy and $N_{0}=\sigma^{2}/\Delta f=\sigma^{2}\tau_{cor}$ is the spectral density of noise, $\Delta f$ is the bandwidth of noise, $\tau_{cor}$ is the noise correlation interval. We see that the probability $P_{\det }$ depends only on the parameter $q=E/N_{0}=\frac{(AMP_{i}^{2})\times W_{i}}{\sigma ^{2}\tau _{cor}}$.
If the energy  is the same both for broad and narrow pulses then $P_{\det }$ will also remain unchanged. For example,  the propagation effects result in broadening the pulse from $W_{i}$ to $mW_{i}$ and the amplitude is reduced to $AMP_{i}/\sqrt{m}$ (to keep  energy $E$ constant) and then we get $\frac{(AMP_{i}^{2}/m)\times mW_{i}}{\sigma ^{2}\iota _{cor}}=q$.\\
Taking this into consideration, it can be conjectured that the cusum algorithm may detect a pulse broadened   due to  moderate dispersion. Only detection is considered  here. Exact reproduction of the transients's  form, of course, requires the knowledge of DM and de-dispersion.\\
However,  conventional practice in the search for transients includes de-dispersion  before time integration.

\begin{figure}
 \includegraphics[width=84mm, height=60mm]{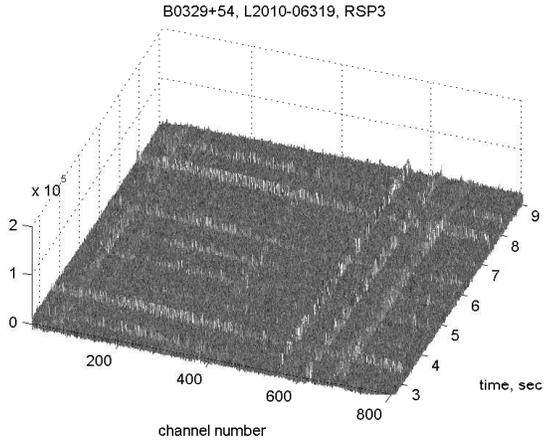}
 \caption{Three-dimensional presentation of the pulsar B0329+54 impulses, LOFAR observational data, 800 frequency channels,  left channel frequency  is 
 $f_{0}=175.29296875MHz$,  channel's bandwidth is $\delta f=0.01220703125MHz$,  time sample interval is $\delta t=0.00131072s$. RFI are also visible.}

 \label{KS-5}
 \end{figure}

\begin{figure}
 \includegraphics[width=84mm, height=60mm]{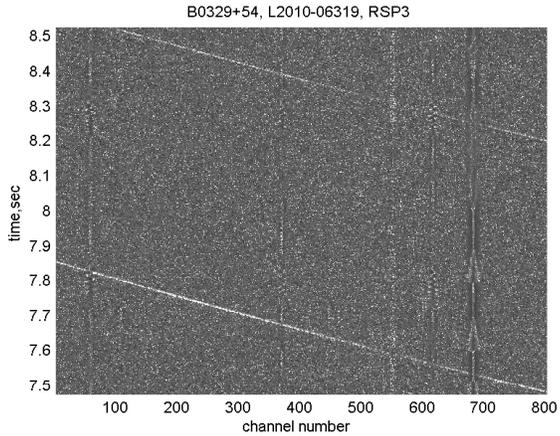}
 \caption{Pulsar tracks from Fig. \ref{KS-5} chosen for the dispersion measure analysis.}
 \label{KS-6}
 \end{figure}

\begin{figure}
  \includegraphics[width=84mm, height=60mm]{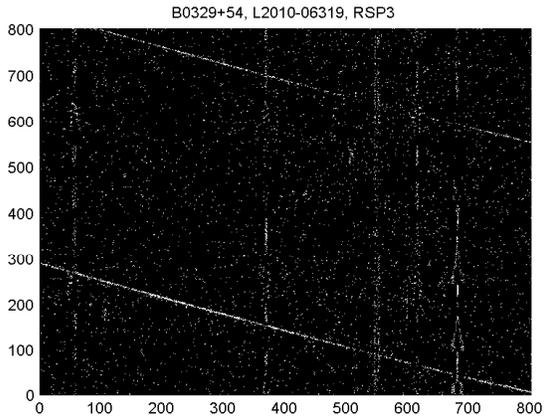}
 \caption{Binary $800 \times 800$ image corresponding to Fig. \ref{KS-6}.}
 \label{KS-7}
 \end{figure}

\begin{figure}
 \includegraphics[width=84mm, height=60mm]{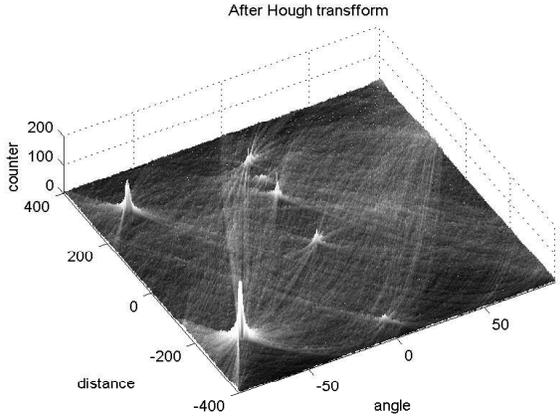}
 \caption{Hough transform image of Fig. \ref{KS-7}.}
 \label{KS-8}
 \end{figure}

 Here we consider another  effective method of non-coherent de-dispersion:  the application of the Hough transform (HT) which is widely used in image processing for line detection, see \cite{duda} and the survey of \cite{Illin}. A transient's track without de-dispersion on the time-frequency $\tau - f$ plane is the second order curve which   can often be approximated by a straight line if the  bandwidth is sufficiently small.  Here we show how HT can be used to detect a straight line on the
time-frequency $\tau - f$ plane. Details of HT and example of computer simulaten are given in Appendix A.\\
 The following example will  demonstrate the  application of HT  to  LOFAR observational data.  Fig. \ref{KS-5} represents the three-dimensional image of the $\tau - f$ plane with the pulsar B0329+54 impulses,  800 frequency channels, each channel's bandwidth is $\delta f=0.01220703125MHz$,  time sample interval $\delta t=0.00131072s$. The left channel frequency is $f_{0}=175.29296875MHz$. Several RFI are also visible on the left and right side of the image.\\
The $800 \times 800$ pixels fragment shown in Fig. \ref{KS-6} around the 8.0s area is chosen for  processing in the following way. The threshold equal to $\approx mean + rms$ of the data's values is applied to the array of numbers represented in Fig. \ref{KS-6} and the corresponding binary image is shown in Fig. \ref{KS-7}. This binary image is then Hough-transformed and the result is shown in Fig. \ref{KS-8}. \\
There are two distinct peaks on the left side of this figure corresponding to the two  pulsar's tracks in Fig. \ref{KS-6} and Fig. \ref{KS-7}. The angular coordinate $\theta$ of these peaks which is the slope of the pulsar's tracks is the same and is equal to $\theta=-69.75^{\circ}$. The module of this value can be recalculated to the dispersion measure: $DM=26.8$ which is not far from the tabulated value for this pulsar ($26.7$). The distances from the center of the initial  image to the straight lines are measured by another coordinate $\rho$. So, HT gives the value of DM  which can be used in delay alignment between frequency channels, similar to    non-coherent de-dispersion.\\
There are  several other peaks in the center of  Fig. \ref{KS-8} corresponding to RFI, but their angle coordinates  $\theta \approx 0$ because the tracks of these narrow-band RFI  are perpendicular to the frequency axis (no dispersion). This is a convenient way to distinguish the  transient's  tracks with nonzero DM from RFI on the $\tau - f$ plane.


When a transient's track on the $\tau - f$ plane cannot be approximated by the straight line, HT can be modified to detect the second order curve. There are many versions of HT adapted to detection of particular curves, see \cite{rao}.

The additional useful property of the HT is  the accumulating of  a number of  points belonging to the same straight line, which is, in essence,  averaging.
Having the slope of the track of a transient on the $\tau - f$ plane it is not only possible  to estimate its dispersion measure with HT , but  also to detect  a weak transient using this averaging property of HT.
 The computer simulation example in Appendix A shows that even for a weak signal (a line on the time-frequency plane) the distinct peak can be detected after HT.

 The cusum algorithm can also be applied to  data on the $\rho - \theta$ Hough plane. Tracks belonging to the same transient form the set of parallel adjacent lines or the strip. The width of the strip is  proportional to the duration of the transient. After HT this strip will be represented by the vertical set of adjacent points on the  $\rho - \theta$ plane: coordinate $\theta$ is  determined by the  slope of the strip and the  range of coordinate $\rho$ is proportional to the width of the strip or the duration of the transient. Therefore, cusum can be calculated along the axis $\rho$  for each  $\theta$ in the same way as it is made on the $\tau - f$ plane and weak transients will be better detected.

\section{Statistical tests after  change detection}
\begin{figure}
 \includegraphics[width=84mm, height=60mm]{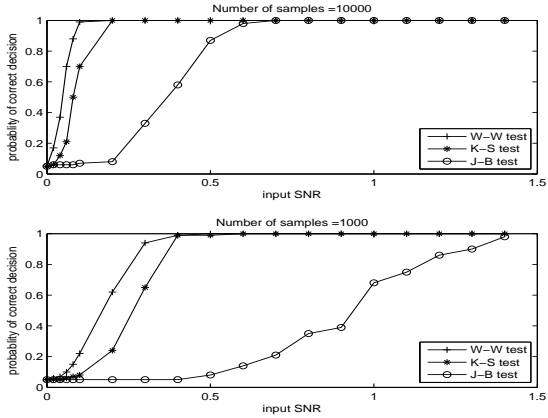}
 \caption{Results of statistical testing of data (signal or RFI) for  methods:
  Wald-Wolfowitz (runs test, W-W), Kolmogorov-Smirnov (K-S) and Jarque–Bera (J-B). Upper panel: the number of samples $M=10^{4}$, lower panel: the  number of samples $M=10^{3}$.}
 \label{table1}
 \end{figure}

The detected signal may be of natural (celestial) origin or  be RFI. To distinguish between these two hypotheses some statistical tests can be applied to the input data. Three of them were studied here:\\
1. The Wald-Wolfowitz  (W-W) test (runs test) which checks the  randomness of the data \citep{Bradley}; \\
2. The single-sample Kolmogorov-Smirnov (K - S) test of the goodness-of-fit  to the normal cumulative distribution function (cdf)\citep{Eadie};\\
3. The Jarque\&Bera (J-B)  goodness-of-fit test using  sample skewness and  sample kurtosis \citep{Jarque}.\\
 Stationarity of the random process ``inside'' a transient is assumed. The test data were divided into 2 groups: the first group - noise with   normal pdf ${\cal N} (0,\sigma_{sys})$, and the second group - noise with  normal pdf ${\cal N} (0,\sigma_{sys})$ plus BPSK (binary phase-shift keying) signal which was modelled by $s_{i}$=$amp\times \sin (2\pi Fi+\phi_{i}),\phi_{i}=\phi_{i-1}+\pi \times psn(\lambda)$, where
$F=0.1$ and $psn(\lambda)$ is the random value with Poisson pdf, parameter $\lambda = 0.1$. The amplitude $amp$ is the parameter in the following tests. All three statistical tests were tuned on the significance level (false alarm probability) $\alpha=0.05$. Each test was repeated $N_{1}=100$ times with the data (number of samples $M=10^{4}$ and $M=10^{3}$) corresponding to
$amp=0$, no change, and then $N_{2}=100$ times  for $\sigma_{s}^{2}>0$ (noise-like signal) or   $amp>0$ (RFI), i.e., in the case of change. \\
The results  of the first part (system noise +  noise-like signal  with  normal pdf) are similar for all three tests: the probabilities of  a wrong decision
(the signal is non-random or non-Gaussian) are at the level of 0.05, i.e., at the level of false alarms. \\
The results of the second part of testing  are shown in Fig. \ref{table1}. The upper figure shows the probabilities of correct decision as functions of the input RFI (`signal')-to-noise-ratio
$snr=(amp^{2}/2)/ \sigma _{sys}^{2}$  for three tests. The number of samples is equal to $M=10^{4}$.\\
The lower figure shows the probabilities of correct decision as functions of the input RFI (`signal')-to-noise-ratio
  for these tests when the number of samples is equal to $M=10^{3}$.

The main conclusion is that the runs test and Kolmogorov-Smirnov test detect signals with  non-normal pdf earlier than the Jarque\&Bera test. The K-S test uses
the whole empirical pdf for comparison with the normal pdf, whereas the J-B test uses empirical kurtosis (the modeled RFI does not alter the skewness), which has considerable  sample variance:  $\approx 24/M$ for  normal pdf. Therefore,  weak non-Gaussian transients can be better detected by W-W and K-S tests.

The Kolmogorov-Smirnov test was also applied in the case of exponential pdf which is the pdf of the power spectrum at each frequency in the absence of RFI.
Any significant deviation from  exponential pdf at a particular frequency indicates that  the signal at this frequency does  not  correspond to  pure noise with a normal pdf \citep{fridman2}, i.e., can contain RFI. Table 1 shows the results of statistical tests similar to those shown in Fig. \ref{table1}, but only for the K-S test.





\begin{table}
\caption[]{Results of statistical testing of  data (signal or RFI) for the Kolmogorov-Smirnov test in the case of exponential pdf for two numbers of samples $M=10^{4}$ and $M=10^{3}$.}
\centering
\begin{tabular}{c|c|c}
 \hline
snr & $M=10^{4}$ &$M=10^{3}$\\
\hline
0.00 & 0.05 & 0.05\\
0.01 & 0.17 & 0.14\\
0.02 & 0.49 & 0.16\\
0.03 & 0.78 & 0.20\\
0.04 & 0.92 & 0.24\\
0.05 & 1.0 & 0.30\\
0.06 & 1.0 & 0.44\\
0.07 & 1.0 & 0.62\\
0.08 & 1.0 & 0.69\\
0.09 & 1.0 & 0.77\\
0.10 & 1.0 & 0.81\\
0.15 & 1.0 & 0.97\\
0.20 & 1.0 & 1.0\\
\end{tabular}
\end{table}

\begin{figure}
 \includegraphics[width=84mm, height=60mm]{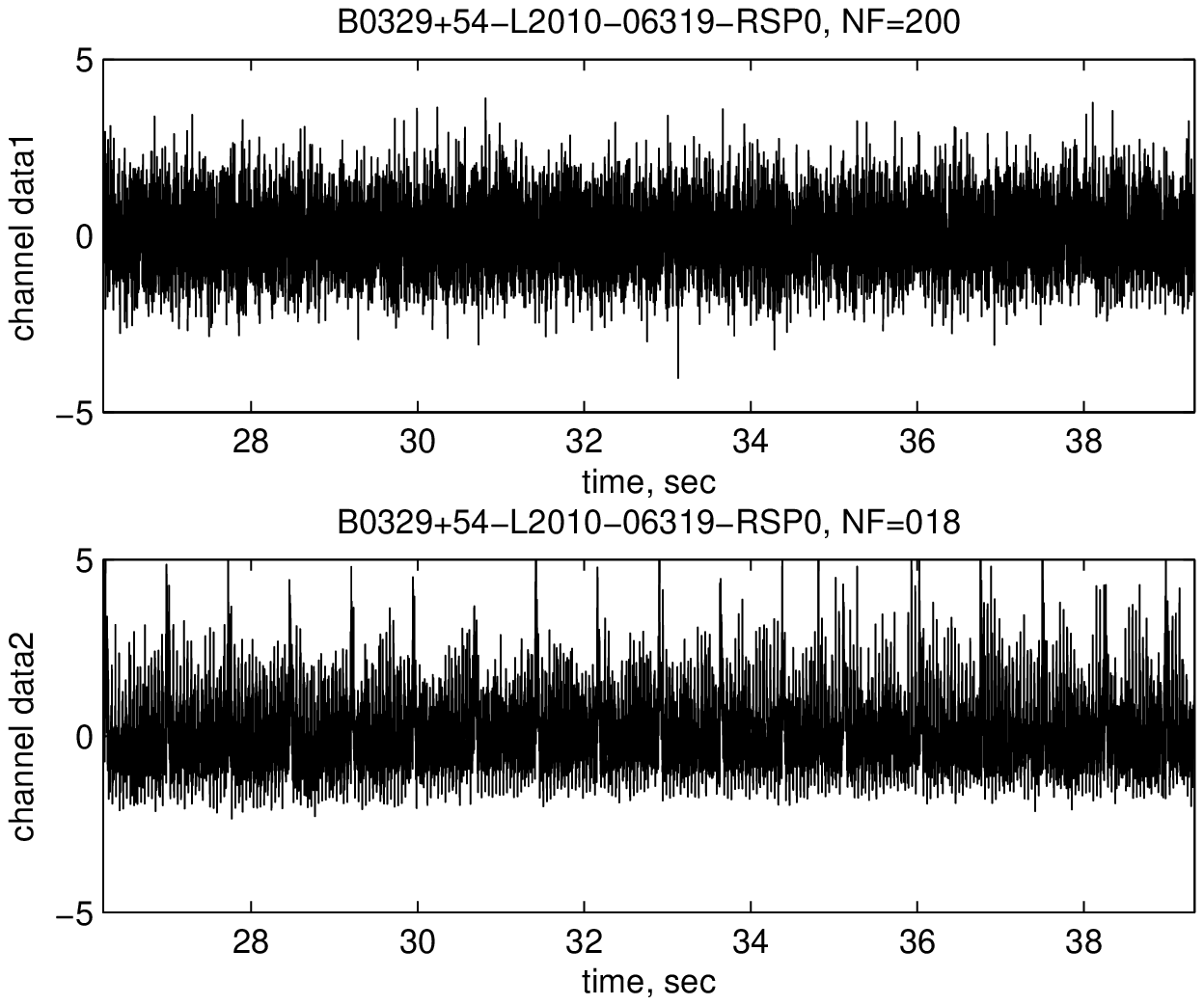}
 \caption{Waveforms of  noise in the reference channel (upper panel) and in the channel with RFI (lower panel).}
 \label{KS-1}
 \end{figure}
\begin{figure}
 \includegraphics[width=84mm, height=60mm]{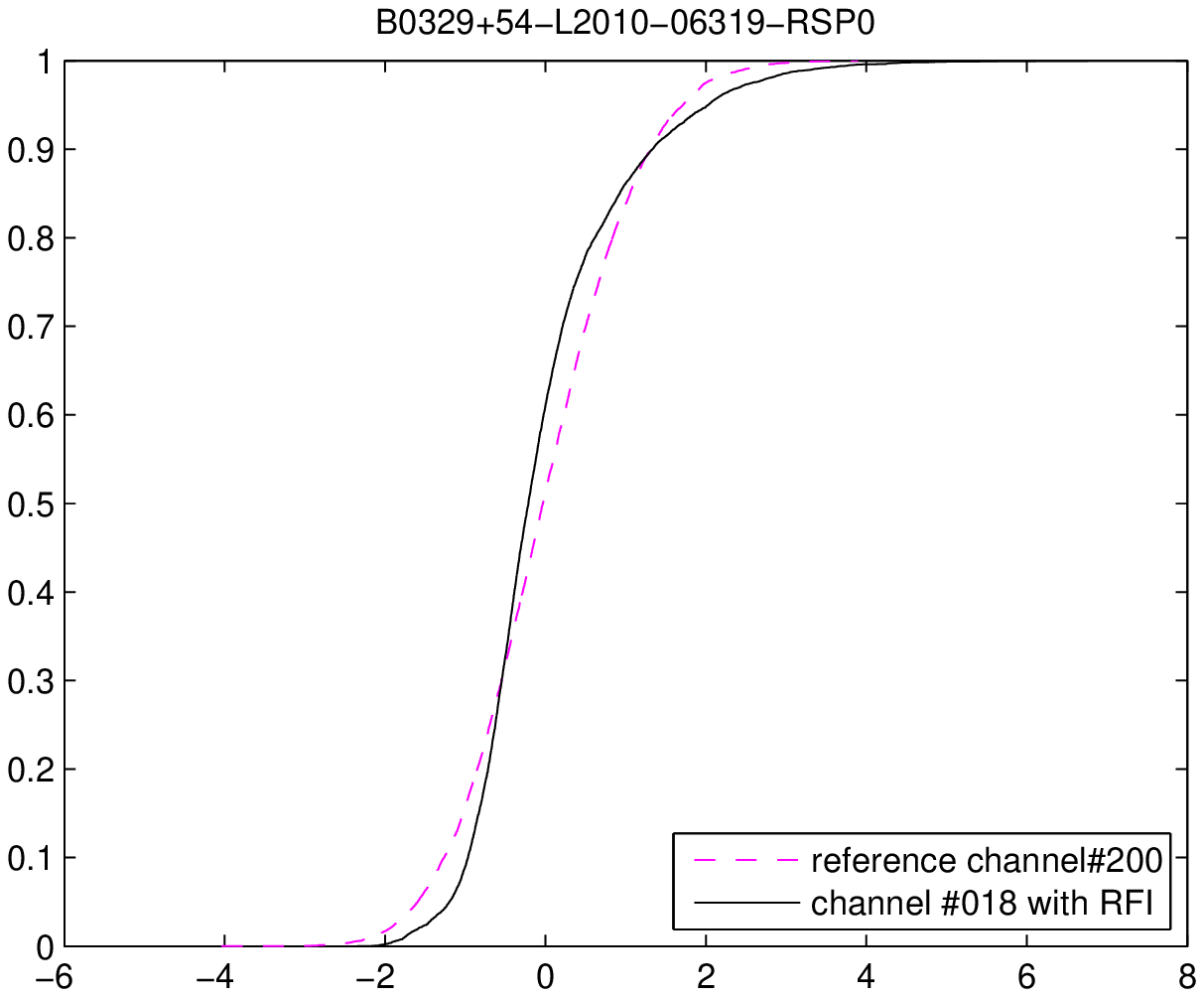}
 \caption{Empirical  cumulative distribution functions of noise in Fig. \ref{KS-1}: in the reference channel (dash line) and in the channel with RFI (solid line).}
 \label{KS-2}
 \end{figure}

 \begin{figure}
 \includegraphics[width=84mm, height=60mm]{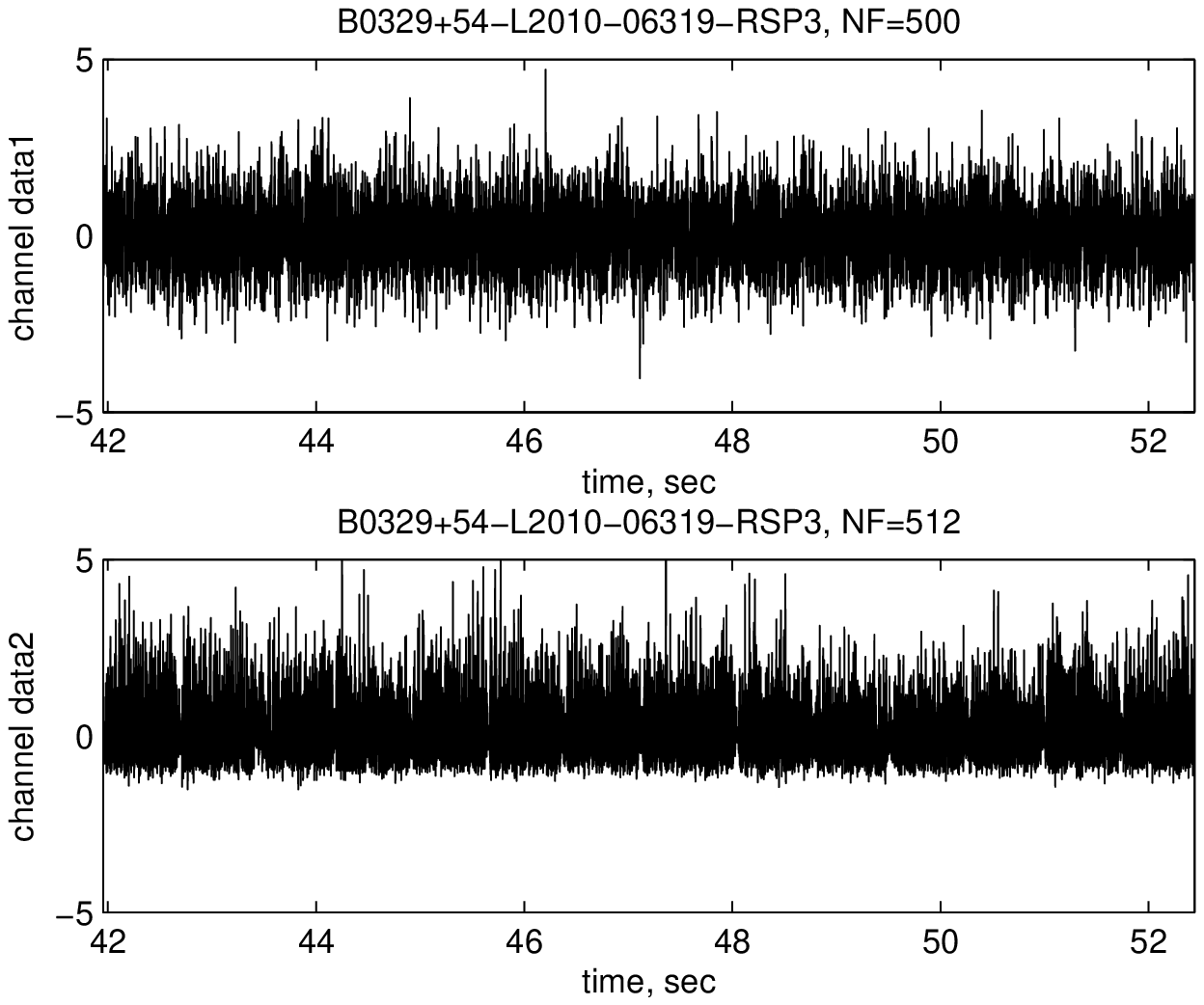}
 \caption{Waveforms of  noise in the reference channel (upper panel) and in the channel with RFI (lower panel).}
 \label{KS-3}
 \end{figure}
\begin{figure}
 \includegraphics[width=84mm, height=60mm]{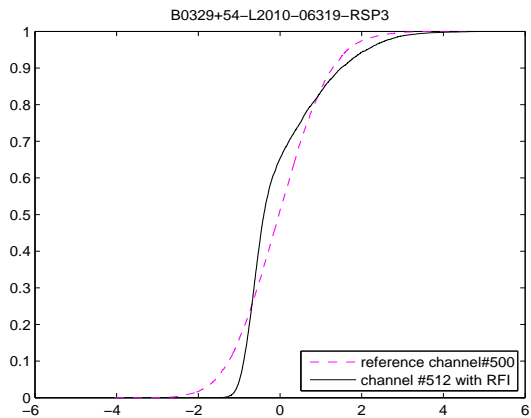}
 \caption{Empirical  cumulative distribution functions of noise in Fig. \ref{KS-3}: in the reference channel (dash line) and in the channel with RFI (solid line).}
 \label{KS-4}
 \end{figure}

A comment must be made here. In practice, even  system  noise  rarely has ideal normal pdf. There are always  slight deviations due to digitization  and digital filtering. The deviations which are visible in histograms are also easily detected by  statistical tests. This  pdf is  usually  stable enough to be used as a reference function for  the two-sample K-S test, etc., and  changes in  the distribution function produced by RFI must be compared with this reference  system noise non-normal pdf.

Figures \ref{KS-1} - \ref{KS-4}  illustrate the application of these statistical tests to  real observational data. Signals at the LOFAR pulsar backend were used. There are four  992-channel filter banks, each frequency channel having the band-width equal to $\approx 12.2KHz $. Some of these  channels contain RFI. The channel $\#029$ (the central frequency=139.3066 MHz) without RFI was chosen as the reference channel (see the waveform of $10^{4}$ samples of noise in this channel in Fig. \ref{KS-1}, the upper panel). The waveform of the channel $\#018$ (the central frequency=139.1724 MHz) with RFI is shown in the lower panel of Fig. \ref{KS-1}.
All above-mentioned tests show  non-Gaussian and non-random (runs test) behaviour of  noise in  channel $\#018$. \\
The two-sample Kolmogorov-Smirnov test was also applied to distinguish the difference in the empirical  pdf of the noise in the reference channel and in the channel with RFI. The Gaussian property was not tested here, only the  deviation from the reference pdf. Fig. \ref{KS-2} shows these two empirical pdf. The distinction between  the two curves is clearly visible.\\
Figures \ref{KS-3} and  \ref{KS-4}  show another example: the reference channel being $\#500$, (the central frequency=181.3965 MHz)   and the channel with RFI $\#512$, (the central frequency=181.5430 MHz). In this case too all tests indicate the presence of RFI.

 The RFI detection algorithms described in this section  must be  "switched on"  {\it after} the  detection of a transient for the purpose of diagnostics, i. e., to reject RFI transients. So, the computational burden is not very large. The computational complexity for W-W and K-S  is proportional to $nlog(n)$, $n$ is the number  of  samples. The  J-B test includes calculation of statistical moments and the  time of completion is proportional to $n$.\\
  All  known and relevant methods of RFI mitigation can be used during observations
but I have here considered only the case of when a transient has been detected and it is necessary to distinguish it from RFI.

\section{Conclusions}

1. Searches in several dimensions (time, frequency, dispersion measure) are a  considerable computational burden during transients detection observations and processing. The cumulative sum method which is the  modification of  Wald's sequential analysis can be helpful in this situation. Computer simulation and the processing of  real observational data show that the time-consuming search into the duration of  a transient (duration-matching) can be reduced using the cusum algorithm. Cusum can be used both after coherent and non-coherent de-dispersion.   The number of computer instructions required by cusum grows linearly  with the number of data samples $n$. The benefit of using cusum is proportional to the number of trials which would be undertaken with the  conventional iterative matching procedure. One of the trial procedures in the search for transients can  therefore be  eliminated.

2.The number of trials  during non-coherent de-dispersion can be reduced using the Hough transform which measures the slopes of the  transient's tracks on the time-frequency plane and, as a result, gives estimates of DM. The averaging property of HT  contributes to the detection of  weak transients. The combination of cusum algorithm and HT provides both an estimation of DM and a duration matching.\\
Due to parallelism and the ``binary'' character of HT many fast HT algorithms were developed in recent years to be implemented in FPGA and multi-core computing systems. Cusum can also be easily mapped on FPGA. In the situation of the {\it a priori} uncertainty both of DM and duration this property promises to be useful, including in  {\it real-time} searches for transients.

3.  Cusum is also a good tool for RFI detection \citep{fridman1}. Burst-like RFI,  erroneously identified as  natural transients, can be distinguished from the noise-like signal-of-interest by the difference in statistics: as a rule, RFI are non-random and non-Gaussian. The proposed statistical tests  allow RFI and transients of natural origin to be  distinguished from one another. The tests  are based on the statistical analysis of the random numbers representing a transient. They are recommended to be used after the detection of a transient and require moderate computational efforts: only  data suspected as being that of a transient are analyzed and the number of computer instructions is proportional to $\approx nlog(n)$.
\\There are, of course, many other methods of RFI mitigation which can be applied  during the entire  observation in order to prevent  false alarms. Everything depends on the type of RFI which is hindering   observations.

4. The methods proposed in this article  do not exclude the use of the  most powerful likelihood criterion of the  transient's celestial origin:  simultaneous observations of the transient signal  coming from the same area of the sky to several radio telescopes which are far distant from each other.

\section{Acknowledgements}

I am grateful to Ben Stappers and Jason Hessels  for the observational data they made available to me for use in this paper.

\appendix

\section[]{The Hough Transform as a tool for the detection of a dispersed transient}

\begin{figure}
 \includegraphics[width=84mm, height=60mm]{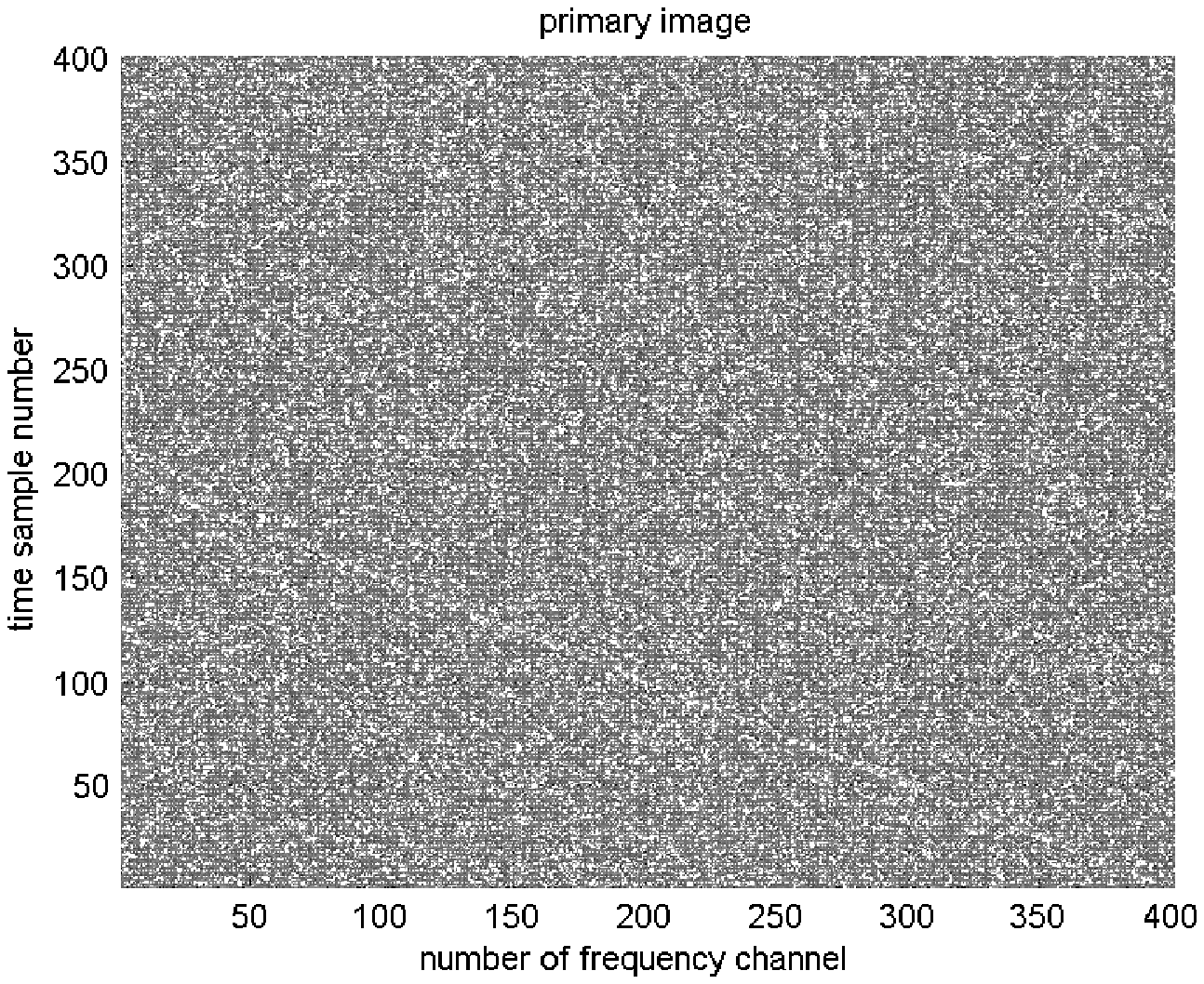}
 \caption{Image of  Gaussian noise with zero mean and $rms=\sigma=1.0$ and  superposed straight line $y=ax+b$, $a=-2.0$ and $b=400$. The amplitude of the line is $A_{T}=1.0$.}
 \label{KS-10}
 \end{figure}

\begin{figure}
 \includegraphics[width=84mm, height=60mm]{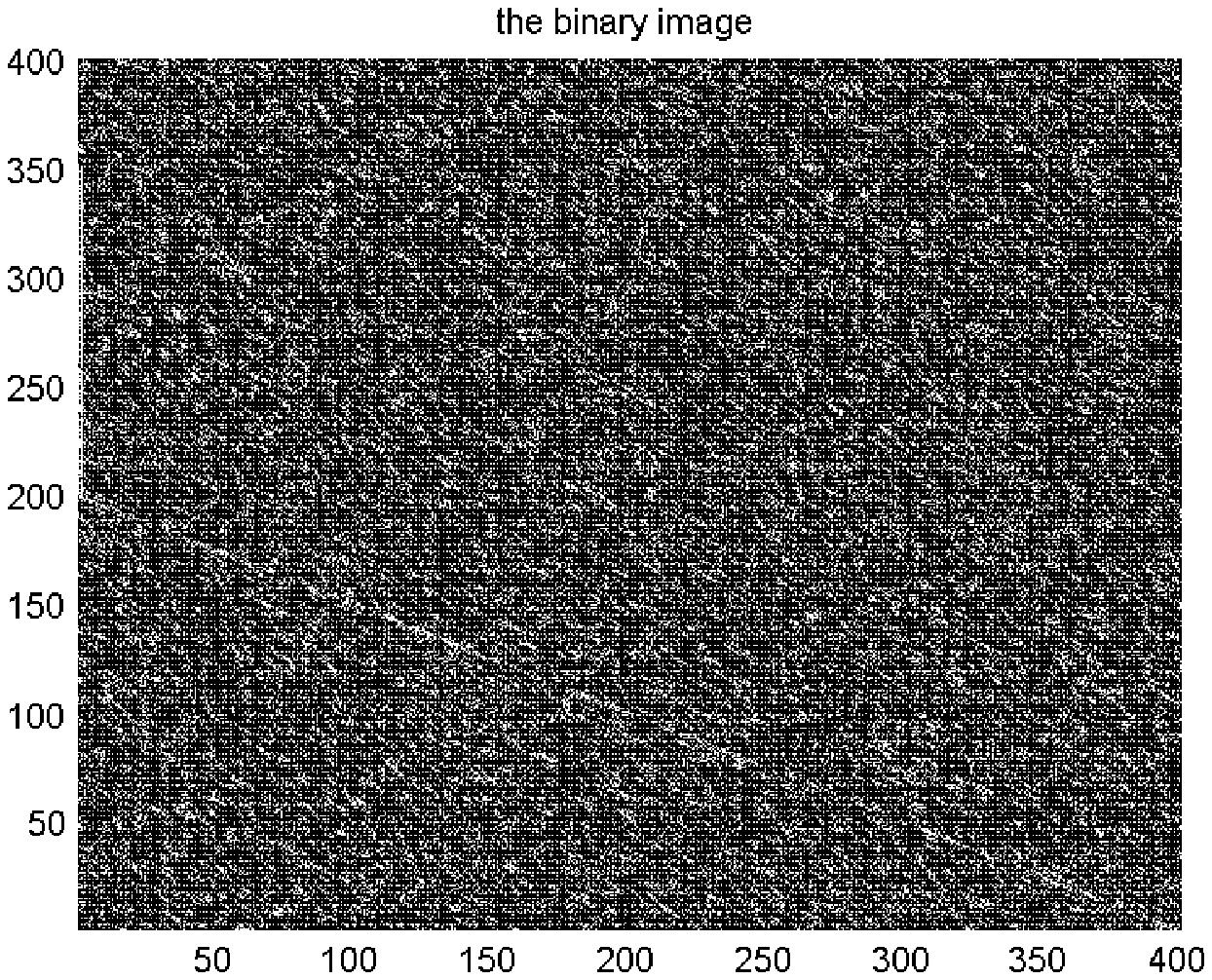}
 \caption{Binary version of image in Fig. \ref{KS-10},   threshold  $thr=\sigma$.}
 \label{KS-11}
 \end{figure}

\begin{figure}
 \includegraphics[width=84mm, height=60mm]{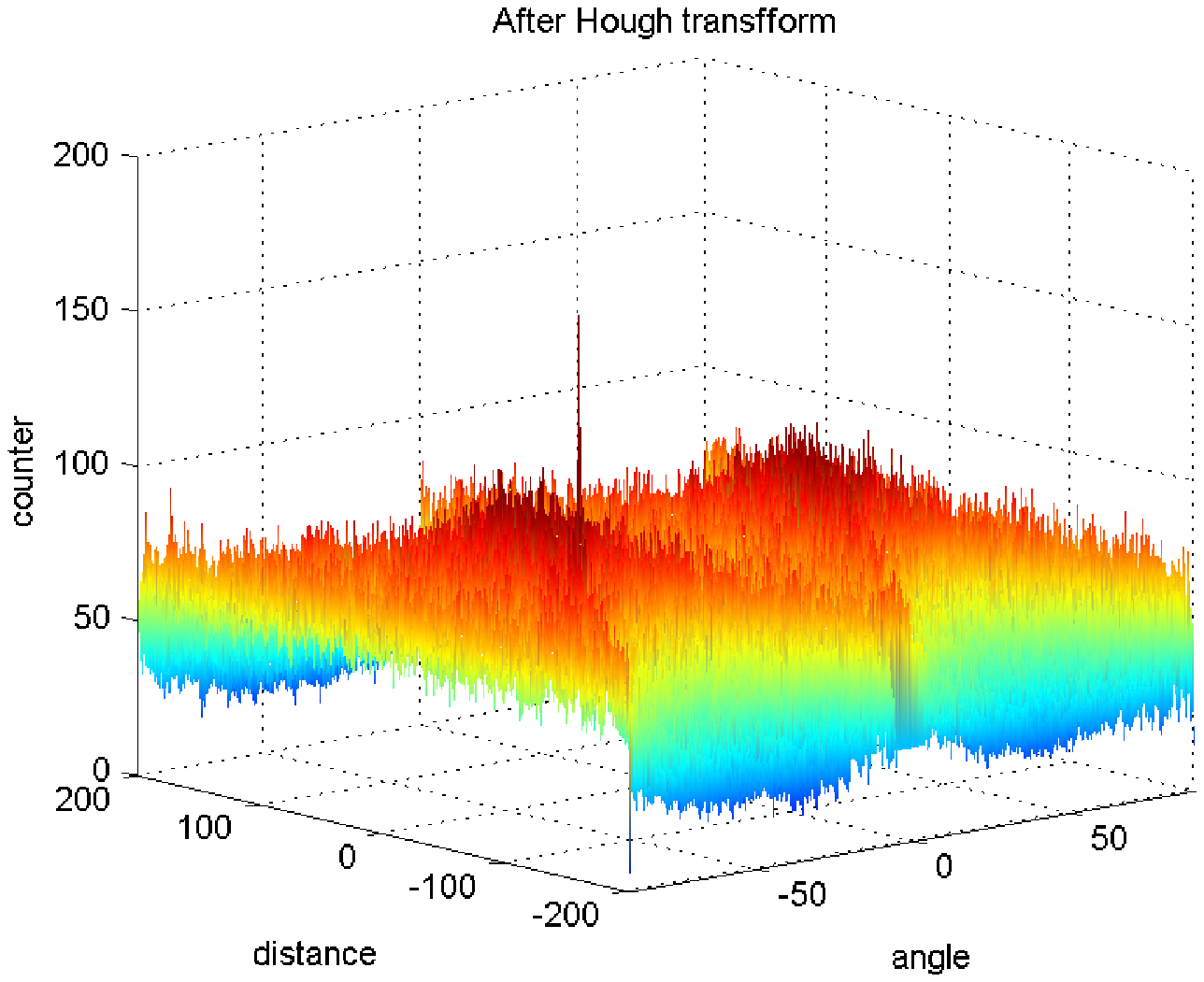}
 \caption{Three-dimensional presentation of  the Hough transform of the image in  Fig. \ref{KS-11}.
 Coordinates $[0,0]$ are in the centre of the image. Peak corresponding to the straight line is  at $\theta=-63.22^{\circ}$ and $\rho=-90$. }
 \label{KS-12}
 \end{figure}

\begin{figure}
 \includegraphics[width=84mm, height=60mm]{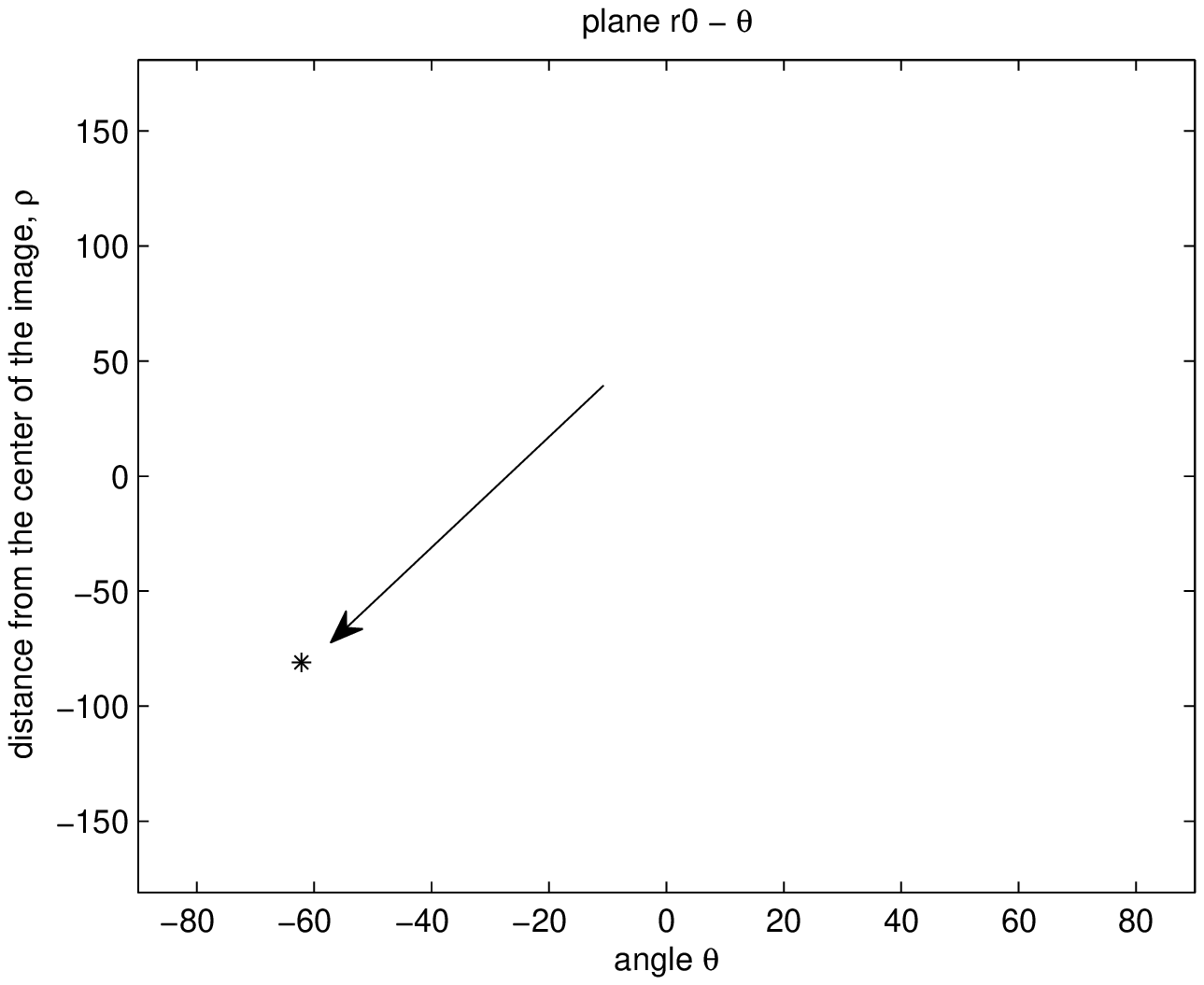}
 \caption{The $(\rho, \theta)$ plane of the Hough transform of the image in  Fig. \ref{KS-11}.
 Coordinates $[0,0]$ are in the centre of the image. The peak indicated by the arrow is  at $\theta=-63.22^{\circ}$ and $\rho=-90$. Precise value of  $\theta=-63.43^{\circ}$}
 \label{KS-13}
 \end{figure}

The Hough Transform (HT) is a useful algorithm for  straight line detection in  binary images  when amplitudes of pixels are equal to two numbers, for example, 1 or 0.
Each straight line $y=ax+b$ on a plane can be parameterized by the angle $\theta$ of its normal to the horizontal axis and its distance $\rho$ from the origin of coordinates. The equation of a line is
\begin{equation}
y=-x\frac{\cos (\theta )}{\sin (\theta )}+\frac{\rho }{\sin (\theta )}.
\end{equation}
This equation can be rewritten for  $\rho$ as a function of $(x,y)$:
\begin{equation}
\rho=xcos(\theta)+ysin(\theta).
\end{equation}
A line can
then be transformed into a single point in the parameter space $(\rho, \theta)$ which  is
called the Hough space. For any pixel in the image with  position $(x, y)$, an infinite number of lines can go
through that single pixel. By using equation (A2) all  pixels belonging to the line can be transformed into the Hough space.
A pixel is transformed into a sinusoidal curve that is unique for this pixel. Doing the same transformation for another pixel gives
another curve that intersects the first curve at one point in the Hough space. This point represents the straight
line  in the image space that goes through both pixels. This operation is repeated for all  pixels of the image.
The pixels belonging to the same straight line have the same point of intersection in the Hough space.\\
 For each point in the Hough space the HT program assigns a  counter which accumulates the number of these intersections. So if there is a straight line with the parameters $(\rho_{i}, \theta_{j})$ consisting of $N$ pixels the counter corresponding to the point $(i,j)$ in the Hough space will contain number $N$.
 This is correct only for the ideal case when there is no noise in the image. In the case of a noisy image the situation is the following. \\
 Let the time-frequency $\tau -  f$ plane being the image on which the search for transients is performed. Each line of pixels along the time axis represents  ``intensity'' samples at the i-th output of FFT or a filter bank.  \\
  In the absence of  transients, these samples  are random numbers  - often with a  Gaussian pdf due to preliminary averaging. A binary image is created  using the threshold $thr$ equal to
 $thr \approx \sigma+m$
  where $\sigma$ is the {\it rms} of the noise samples and $m$ is the mean value. If the amplitude of a sample is less than $thr$ it is converted to 0, otherwise it is equal to 1. Now the binary image is covered with the random 0 and 1 and the probability $p$ of 1 is equal to $\approx 0.16$. For the size of the image $M \times M$ the number of 1's is $\approx p\times M^{2}$. It is difficult to  calculate precisely  noise pdf on the $(\rho, \theta)$ plane. So  the approximate estimates of the first two moments, confirmed by computer simulation, are given here: mean$\approx p\times M$, rms$\approx \sqrt{p(1-p)M}$.\\
 The image on the $\tau -  f$ plane of a transient after  dispersion due to wave propagation in the interstellar media  is a second-order curve line. In a narrow range of frequencies the curve can be approximated as a straight line. The width of the  band for which this approximation is valid depends on the observational sky frequency.\\
  To better understand  the HT  a computer simulation example is given here. Let the  $400 \times 400$ image in Fig. \ref {KS-10} represent the ``noisy'' $\tau -  f$ plane with zero mean and $\sigma=1.0$. There is also the straight line imitating  the track of a dispersed transient. Equation of this line in the coordinates  $(x,y)$ is  $y=ax+b$ where $a=-2.0$ and $b=400$. The amplitude of the line is $A_{T}=1.0$. After the threshold $thr=\sigma$  we get the  binary image  shown in Fig. \ref{KS-11}. The Hough transform of the image in Fig. \ref{KS-11} is represented in Fig. \ref{KS-12} where the peak corresponds to the straight line in Fig. \ref {KS-10}. The  search for the maximum amplitude of the peak gives the estimate $\widehat{\theta }=-63.22^{\circ}$ while the exact number is $\theta=-63.43^{\circ}$, see Fig. \ref{KS-13}. The origin of coordinates in the Hough plane is positioned at the center of the image.

 The useful property of the HT is  in the accumulation of the number $N$ of coincident points in each cell of the Hough image (belonging to the same straight line),  thus  averaging samples along the line on the primary image.
 If the initial $SNR=A_{T}/ \sigma \approx1.0$,  as in our example, the expected $SNR$ in the Hough plane will be $\approx \sqrt{N}$.

 Therefore, it is not only possible  to estimate the dispersion measure with the help of HT by the slope of a transient's track, but also to detect  a weak transient, using the averaging property of HT.

\begin{figure}
 \includegraphics[width=84mm, height=60mm]{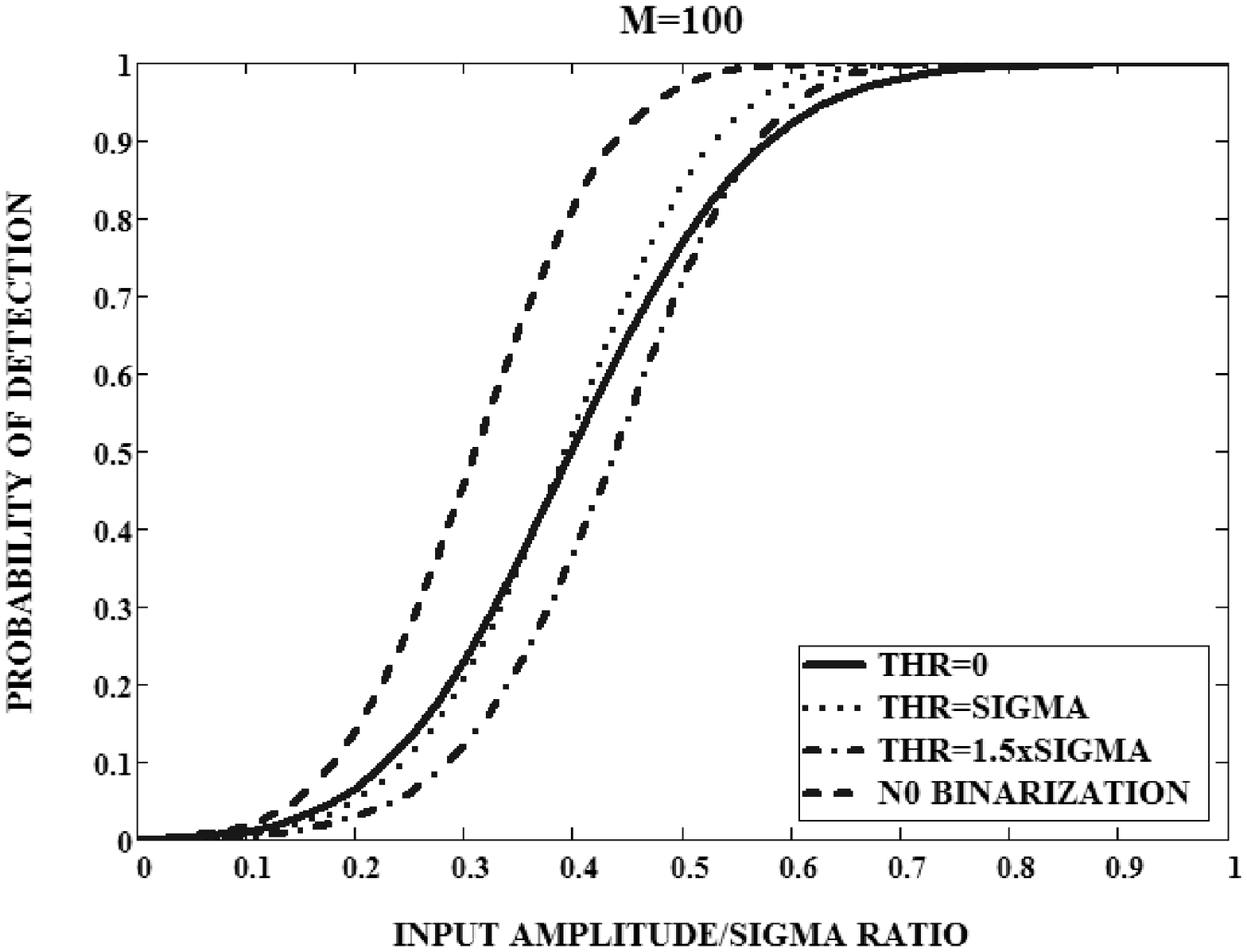}
 \caption{Probability of detection after averaging along the transient track, the number of samples in the track $M=100$: without binarization and binarization with different thresholds $thr=0;\sigma;1.5\sigma$. The probability of false alarms $\alpha=10^{-3}$.}
 \label{KS-14}
 \end{figure}

  Transition from the primary image ($\tau - f$ plane) to the binary image leads to some loss. It is convenient to estimate this loss by comparing the probability of detection $P_{det}$ after averaging along the transient track with and without binarization. Fig.  \ref {KS-14} shows these $P_{det}$ as functions of the normalized amplitude $A_{T}/\sigma$ of the track for four cases: without binarization and binarization with three different thresholds
 $thr=0;\sigma;1.5\sigma$. The number of averaged samples along the track is $M=100$. The probability of false alarms for all four curves is $\alpha=10^{-3}$.\\
 Comparing these curves the following conclusions can be made:\\
 a) there is a certain loss ($\approx 1.5dB$) due to binarization;\\
 b) there is no practical differences between the curves with binarization in the area $P_{det}\approx 1.0$.

\label{lastpage}
\end{document}